\documentclass[9t,letterpaper]{article}
\usepackage{authblk}
\usepackage{lscape}
\usepackage{ifluatex}
\ifluatex
  \usepackage{fontspec}
\fi
\usepackage{tabularx,booktabs}
\usepackage{footnote}
\usepackage{amssymb}
\usepackage{amsmath}
\usepackage{bbold}
\usepackage{enumerate}
\usepackage{datetime}

\usepackage{url}
\usepackage[dvipsnames,svgnames]{xcolor}

\makeatletter

 \let\proof\@undefined
 \let\endproof\@undefined
\makeatother

\usepackage{amsthm}
\newtheoremstyle{ssstyle}                
  {.5em}                                 
  {.5em}                                 
  {\rmfamily\slshape}                    
  {0em}                                  
  {\rmfamily\bfseries}                   
  {.}                                    
  { }                                    
  {\thmname{#1}\thmnumber{ #2} (\thmnote{#3})} 
\newtheoremstyle{ssthmstyle}                
  {.5em}                                 
  {.5em}                                 
  {\rmfamily\slshape}                    
  {0em}                                  
  {\rmfamily\bfseries}                   
  {.}                                    
  { }                                    
  {\thmname{#1}\thmnumber{ #2}} 
\theoremstyle{ssstyle}
\newtheorem{ssdef}{Definition}

\theoremstyle{ssthmstyle}

\DeclareMathOperator*{\argmax}{arg\,max}
\DeclareMathOperator*{\argmin}{arg\,min}
\usepackage[ruled, vlined, algo2e, nokwfunc, norelsize]{algorithm2e}
\SetNlSty{}{\sffamily}{}
\SetAlgoNlRelativeSize{-4}
\SetAlCapFnt{\bfseries}

\SetCommentSty{mycommentfn}
\SetKwComment{com}{$\triangleright$ }{}
\usepackage{enumerate}
\usepackage{ifdraft}
\usepackage[labelfont=rm, textfont=rm, caption=false]{subfig}
\usepackage{balance}
\newcommand{\ssparhd}[1]{\medskip\noindent\textbf{#1.}\ }
\usepackage{booktabs}
\usepackage{dcolumn}
\usepackage{rotating}
\usepackage{multirow}
\newcommand{\stepsthead}[2]{\multicolumn{1}{#2}{\small\textsf{\bfseries #1}}}
\usepackage{pgfplotstable}
\usepackage{tabularx}
\usepackage{tikz}
\usetikzlibrary{fit}
\usetikzlibrary{backgrounds}
\usetikzlibrary{calc}
\usetikzlibrary{patterns}
\tikzset{tr/.style={circle, fill=none, draw=black, font=\small, inner sep=0pt, minimum size=5mm}}
\usepackage{pgfplots}
\usepgfplotslibrary{groupplots}
\usetikzlibrary{fit}
\usetikzlibrary{patterns}
\definecolor{MPIIbeige}{cmyk}{0.0,0.0,0.08,0.23}
\definecolor{MPIIblue}{cmyk}{0.22,0.11,0.0,0.36}
\definecolor{MPIIdarkblue}{cmyk}{1.0,0.5,0.0,0.72}
\definecolor{MPIIdarkgreen}{cmyk}{1.0,0.0,0.47,0.30}
\definecolor{MPIIgrey}{cmyk}{0.0,0,0.024,0.09}
\definecolor{MPIIlightgreen}{cmyk}{0.25,0.0,1.0,0.20}
\newcommand{\xtl}[1]{{\footnotesize #1}}
\usepackage{enumerate}
\usetikzlibrary{backgrounds,calc}
\newcommand{\ii}{\ensuremath{\mathcal{I}}}
\newcommand{\rr}{\mathcal{R}}
 
\newcommand{\extendedby}[0]{\tikz[baseline=-3pt] \draw (0,0) -- (0,-1.5pt) -- (4pt,-1.5pt) --
  (4pt,0pt) (2pt,0pt) -- (2pt,1.5pt) -- (6pt,1.5pt) -- (6pt,0pt);}

\newcommand{\sscard}[1]{\vert #1\vert}

\newcommand{\oh}{\ensuremath{\mathcal{O}}}
\newcommand{\finex}{} 
\newcommand{\comment}[1]{}
\def\mycomment#1{}

\def\withcomments{
  \addtolength{\oddsidemargin}{-0.5 in}
  \addtolength{\evensidemargin}{-0.5 in}
  \newcounter{mycommentcounter}
  \def\mycomment##1{
    \refstepcounter{mycommentcounter}
    \ifhmode
    \unskip{
      \dimen1=\baselineskip \divide\dimen1 by 2 %
      \raise\dimen1\llap{{ \tiny -\themycommentcounter-}}}
    \fi
    \marginpar{\renewcommand{\baselinestretch}{0.8}%
      \footnotesize [\themycommentcounter]: \raggedright ##1}
    }
  \date{\framebox{Draft of \today}}
  }
\usepackage{flushend}
\newcommand{\ferrari}{{\textrm{{FERRARI}}}\xspace}

\renewcommand{\baselinestretch}{0.98}
\title{High-Performance Reachability Query Processing under Index Size Restrictions}

\author[1]{Stephan~Seufert}
\author[1]{Avishek~Anand}
\author[2]{Srikanta~J.~Bedathur}
\author[1]{Gerhard~Weikum}
\affil[1]{Max Planck Institute for Informatics, Germany \authorcr \vspace*{-4pt}{\small
    \texttt{$\{$sseufert$\vert$aanand$\vert$weikum$\}$@mpi-inf.mpg.de}}\vspace*{4mm}}

\affil[2]{IIIT-D, India \authorcr\vspace*{-4pt}
    {\small \texttt{bedathur@iiitd.ac.in}}}
\date{}

\begin{document}
\maketitle
\begin{abstract} 
  In this paper, we propose a scalable and highly efficient index structure for the reachability
  problem over graphs. We build on the well-known node interval labeling scheme where the set of
  vertices reachable from a particular node is compactly encoded as a collection of node identifier
  ranges. We impose an explicit bound on the size of the index and flexibly assign approximate
  reachability ranges to nodes of the graph such that the number of index probes to answer a query
  is minimized. The resulting tunable index structure generates a better range labeling if the space
  budget is increased, thus providing a direct control over the trade off between index size and the
  query processing performance. By using a fast recursive querying method in conjunction with our
  index structure, we show that in practice, reachability queries can be answered in the order of
  microseconds on an off-the-shelf computer -- even for the case of massive-scale real world graphs.
  Our claims are supported by an extensive set of experimental results using a multitude of
  benchmark and real-world web-scale graph datasets.
\end{abstract}

\section{Introduction}
\label{sec:intro}
Reachability queries are a fundamental operation in graph mining and algorithmics and ample work
exists on index support for reachability problems. In this setting, given a directed graph and a
designated source and target node, the task of the index is to determine whether the graph contains
a path from the source to the target.

Computing reachability between nodes is a building block in many kinds of graph analytics, for
example biological and social network analysis, traffic routing, software analysis, and linked data
on the web, to name a few. In addition, a fast reachability index can prove useful for speeding up
the execution of general graph algorithms -- such as shortest path and Steiner tree computations --
via search-space pruning. As an example, Dijkstra's algorithm can be greatly sped up by avoiding the
expansion of vertices that cannot reach the target node.

While the reachability problem is a light-weight task in terms of its asymptotic complexity,
the advent of massive graph structures comprising hundreds of millions of nodes and billions of
edges can render even simple graph operations computationally challenging. It is thus crucial for
reachability indices to provide answers in sublinear or ideally near-constant time. Further
complicating matters, the index structures, which generally reside in main-memory, are expected to
satisfy an upper-bound on the size. In most scenarios, the available space is scarce, ranging from
little more than enough to store the graph itself to a small multiple of its size.

Given their wide applicability, reachability problems have been one of the research foci in graph
processing over recent years. While many proposed index structures can easily handle small to
medium-size graphs comprising hundreds of thousands of nodes
(e.g.,~\cite{Agrawal1989,Chen2008,Cohen2002,Jin2011,Jin2009,Jin2008,Schenkel2004,Trissl2007,vanSchaik2011,Wang2006}),
massive problem instances still remain a challenge to most of them. The only technique that can cope
with web-scale graphs while satisfying the requirements of restricted index size and fast query
processing time, employs guided online search~\cite{Yildirim2010, Yildirim2011}, leading to an index
structure that is competitive in terms of its construction time and storage space consumption, yet
speeds up reachability query answering significantly when compared to a simple DFS/BFS traversal of
the graph. However, it suffers from two major drawbacks. Firstly, given the demanding constraints on
precomputation time, only \emph{basic heuristics} are used during index construction, which in many
cases leads to a suboptimal use of the available space. Secondly and more importantly, while the
majority of reachability queries involving pairs of nodes that are not reachable can be efficiently
answered, the important class of \emph{positive queries} (i.\,e. the cases in which the graph
actually contains a path from the source to target) has to be regarded as a worst-case scenario due
to the need of recursive querying.  This can severely hurt the performance of many practical
applications where positive queries occur frequently.

The reachability index structure we propose in this paper -- coined FERRARI (for Flexible and
Efficient Reachability Range Assignment for gRaph Indexing) -- overcomes the limitations of existing
approaches by \emph{adaptively compressing the transitive closure during its construction}. This
technique enables the efficient computation of an index geared towards minimizing the expected query
processing time given a user-specified constraint on the resulting index size. Our proposed index
supports positive queries efficently and outperforms GRAIL, the best prior method, on this class of
queries by a large margin, while in the vast majority of our experiments also being faster on
randomly generated queries.  To achieve these performance gains, we adopt the idea of representing
the transitive closure of the graph by assigning identifiers to individual nodes and encoding sets
of reachable vertices by intervals, first introduced by Agrawal et al.~\cite{Agrawal1989}. Instead
of materializing the full set of identifier ranges at every node, we adaptively merge adjacent
intervals into fewer yet coarser representations at construction time, whenever a certain space
budget is exceeded. The result is a collection of \emph{exact and approximate} intervals that are
assigned as labels of the nodes in the graph. These labels allow for a guided online search
procedure that can process positive as well as negative reachability queries significantly faster
than previously proposed size-constrained index structures. \\
The interval assignment underlying our approach is based on the solution of an associated interval
cover problem.  Efficient algorithms for computing such a covering structure together with an
optimized guided online search facilitate an efficient and flexible reachability index structure.

In summary, this paper makes the following technical contributions:
\begin{itemize}
\item a space-adaptive index structure for reachability queries based on selective compression of
  the transitive closure using exact and approximate reachability intervals,
\item efficient algorithms for index construction and querying that allow extremely fast
  query processing on web-scale real world graphs, and
\item extensive experiments that demonstrate the superiority of our approach in comparison to the
  best prior method that satisfies index size constraints, GRAIL.
\end{itemize}

The remainder of the paper is organized as follows: In Section \ref{sec:prelim} we introduce
necessary notation and the basic idea of reachability interval labeling. Afterwards, we give a short
introduction to approximate interval indexing in Section \ref{sec:approx}, followed by an in-depth
treatment of our proposed index (Section \ref{sec:ferrari}). An overview over our query processing
algorithm is given in Section~\ref{sec:qp}, followed by the experimental evaluation and concluding
remarks.


 \section{Preliminaries}
\label{sec:prelim}
In the following, let $G=(V,E)$ denote a directed graph with $n := \vert V\vert$ nodes and $m :=
\vert E\vert$ edges. For a node $v$, let
\begin{align}
  \label{eq:11}
  &\mathcal{N}^+(v) := \{ w \in V \mid (v,w) \in E\} \\
  \textrm{and}\quad &\mathcal{N}^-(v) := \{ u \in V \mid (u,v) \in E\}
\end{align}
denote the sets of nodes with an edge coming from and leading to $v$, respectively.

Given nodes $u,v \in V$, we call $v$ \textit{reachable} from $u$, written as $u \sim v$, if $E$
contains a directed path from $u$ to $v$.  Further, we denote the set of vertices reachable from $v
\in V$ in $G$ by $\rr_G(v) := \{ w \in V \ \vert \ v \sim w \}$. We call $\rr_G(v)$ the
\textit{reachable set} of $v$ (we drop the subscript whenever the graph under consideration is clear
from the context). A \textit{reachability query} with source node $u$ and destination node $v$ of a
graph $G$ is expressed as a triple $(G, u, v)$ and answered with a boolean value by a reachability
index.

A pair $(u,v)$ of nodes exhibits \textit{strong reachability} if $u \sim v$ and $v\sim u$, that is,
$u$ and $v$ are mutually reachable in $G$. Note that strong reachability induces an equivalence
relation on the set of nodes. The equivalence classes of this relation, that is, the maximal subsets
$V' \subseteq V$ with $u \sim v$ for all $u,v \in V'$, are called \textit{strongly connected
  components} of $G$. For a node $v \in V$ let $[v]$ denote the strongly connected component that
contains $v$.

\ssparhd{Condensed Graph} We define the \textit{condensed graph} of $G$, denoted as $G_C =
(V_C,E_C)$, as the graph obtained after collapsing the maximal strongly connected components into
``supernodes'', i.\,e.
\begin{align}
  \label{eq:condensed_graph}
   &V_C \,:= \bigl\{ [v] \ \big\vert \ v \in V \bigr\} \\
   \text{and}\quad 
   &E_C := \bigl\{ \bigl([u],[v]\bigr) \ \big\vert \ (u,v) \in E, [u] \ne [v] \bigr\}.
\end{align}
By definition, $G_C$ is a directed acyclic graph (DAG). \\
It is important to note that the reachability queries $(G,u,v)$ and $(G_C,[u],[v])$ are
equivalent. Thus, the existing index structures (including ours) consider only directed acyclic
graphs, that is, create an index over the condensed graph $G_C$.  At query time, the input nodes
$u,v$ are then mapped to their respective strongly connected components, allowing early
termination whenever $[u]=[v]$.

\ssparhd{Tree Cover and Graph Augmentation} For a graph $G$, we define a \textit{tree cover} of $G$,
denoted as $T(G)$, as a directed spanning tree of $G$. If $G$ contains more than one node with no
incoming edges, instead of a tree only a spanning forest can be obtained.

In this case, we augment the graph $G$ by introducing an artificial root node $r$ that is connected
to every node with no incoming edge: 
\begin{equation}
   \label{eq:rootaugmentation}
   G' := \bigl(V \cup \{r\}, E \cup \{ (r,v) \ \vert \ v \in V,\, \mathcal{N}^-(v) = \emptyset \}\bigl).
\end{equation}
Note that this modification has no effect on the reachability relation among the existing nodes of $G$.

\ssparhd{Integer Intervals} For integers $x,y \in \mathbb{N}, x\leq y$, we use the interval $[x,y]$
to represent the set $\{x, x+1,\ldots, y\}$.  Let $I=[a,b]$ and
$J=[p,q]$ denote integer intervals.  We define $\vert I\vert := b - a + 1$ to denote the number of
elements contained in $I$. Further, we call $J$ \emph{subsumed by} $I$, written $J \sqsubseteq I$,
if $J$ corresponds to a subinterval of $I$, i.\,e
\begin{equation}
   \label{eq:13}
   J \sqsubseteq I \quad\Longleftrightarrow\quad a \leq p \leq q \leq b.
 \end{equation}
Further, $J$ is called an \emph{extension} of $I$, denoted $I \extendedby J$, if the start-point but
not the end-point of $J$ is contained in $I$:
\begin{equation}
   \label{eq:14}
   I \extendedby J \quad\Longleftrightarrow\quad a \leq p \leq b < q. 
 \end{equation}

\begin{table}
 \centering
 \begin{tabularx}{.75\columnwidth}{cl}
   \toprule
   \stepsthead{Symbol}{c} & \stepsthead{Description}{l} \\
   \midrule
   $V$ & set of vertices \hspace*{50mm}$\sscard{V} = n$ \\
   $E$ & set of edges  \hspace*{53.4mm}$\sscard{E} = m$ \\
   $\mathcal{N}^{+/-}(v)$ & set of direct successors/predecessors of $v$ \\
   $\rr_G(v)$ & reachable set of node $v$ in $G$ \\
   $[v]$ & strongly connected component of $v$ \\
   $G_C$ & condensed graph (collapsed SCCs) \\
   $G'$ & augmented graph (virtual root node $r$) \\
   $T(G)$ & tree cover of $G$ \\
   $T_v$ & subtree of tree $T$, rooted at $v$ \\
   $\pi(v)$ & post-order number of node $v$ \\
   $\tau(v)$ & topological order number of node $v$ \\
   $I_T(v)$ & reachability interval of $v$ in tree $T$ \\
   $\mathcal{I}(v)$ & set of reachability intervals of $v$ \\
   \bottomrule
 \end{tabularx}
 \caption{Notation}
 \label{table:symbols}
\end{table}
An overview over the symbols used throughout this paper is given in Table \ref{table:symbols}.


\subsection{Interval Indexing}
\label{sec:interval}
In this section, we introduce the concept of node identifier intervals for reachability processing,
first proposed by Agrawal et al.~\cite{Agrawal1989}, which provided the basis of many subsequent
indexing approaches, including our own. The key idea is to assign numeric identifiers to the nodes
in the graph and represent the reachable sets of vertices in a compressed form by means
of interval representations. This technique is based on the construction of a tree cover of the
graph followed by post-order labeling of the vertices.

Let $G'$ denote the augmented input graph as defined above.  Further, let $T=(V_T,E_T)$ denote a
tree cover of $G'$. In order to assign node identifiers, the tree is traversed in depth-first
manner. In this setting, a node $v$ is visited after all its children have been visited. The
post-order number $\pi(v)$ corresponds to the order of $v$ in the sequence of visited nodes.

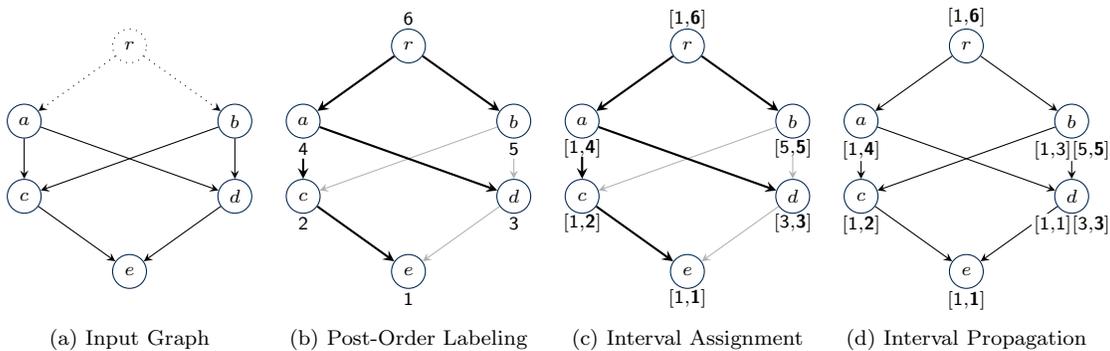
\begin{figure*}
  \centering
  \def\xsc{1.4}
\def\ysc{1}
\subfloat[Input Graph]{
  \begin{tikzpicture}[baseline=-4ex, xscale=\xsc, yscale=\ysc, tr/.style={font=\scriptsize, circle, draw=MPIIdarkblue, minimum size=4.5mm, inner sep=0pt}, >=stealth]
    \foreach \x/\y/\l in {0/1/c, 2/1/d, 0/2/a, 2/2/b, 1/0/e} 
    { 
      \node[tr] (\l) at (\x,\y) {$\l$}; 
    }
    \foreach \s/\t in {b/d, b/c, a/d, a/c, c/e, d/e} 
    {
      \draw[->] (\s) -- (\t);
    }
    \node[tr, dotted] (r) at (1,3) {$r$};
    \path[->, dotted] (r) edge (a) (r) edge (b);
  \end{tikzpicture}
}
\hfill
\subfloat[Post-Order Labeling]{
  \begin{tikzpicture}[baseline=-4ex, xscale=\xsc, yscale=\ysc, tr/.style={font=\scriptsize, circle,
      draw=MPIIdarkblue, minimum size=4.5mm, inner sep=0pt, fill=white}, >=stealth]
    \foreach \x/\y/\l in {1/3/r, 0/1/c, 2/1/d, 0/2/a, 2/2/b, 1/0/e} 
    { 
      \node[tr] (\l) at (\x,\y) {$\l$}; 
    }

    \begin{pgfonlayer}{background}
    \foreach \s/\t in {d/e, b/c, b/d} {
      \draw[->, black!30, thin] (\s) -- (\t);
    }
    \foreach \s/\t in {r/a, a/c, a/d, c/e, r/b} 
    {
      \draw[->, thick] (\s) -- (\t);
    }
    \end{pgfonlayer}
    \begin{pgfonlayer}{background}
      \tikzstyle{label}=[font=\scriptsize, overlay, fill=white, draw=none,
         rounded corners, inner sep=1.5pt, outer sep=0pt, text width=3mm, text centered]
      \foreach \i/\l in {e/1, c/2, a/4, d/3, b/5} 
      {
        \node[label, anchor=north] at (\i.south) (int\i) {$\mathsf{\l}$}; 
      }
      \foreach \i/\l in {r/6} 
      {
        \node[label, anchor=south] at (\i.north) (int\i) {$\mathsf{\l}$}; 
      }
    \end{pgfonlayer}
  \end{tikzpicture}
}
\hfill
\subfloat[Interval Assignment]{
  \begin{tikzpicture}[baseline=-4ex, xscale=\xsc, yscale=\ysc, tr/.style={font=\scriptsize, circle,
      draw=MPIIdarkblue, minimum size=4.5mm, inner sep=0pt, fill=white}, >=stealth]
    \foreach \x/\y/\l in {1/3/r, 0/1/c, 2/1/d, 0/2/a, 2/2/b, 1/0/e} 
    { 
      \node[tr] (\l) at (\x,\y) {$\l$}; 
    }

    \begin{pgfonlayer}{background}
    \foreach \s/\t in {d/e, b/c, b/d} {
      \draw[->, black!30, thin] (\s) -- (\t);
    }
    \foreach \s/\t in {r/a, a/c, a/d, c/e, r/b} 
    {
      \draw[->, thick] (\s) -- (\t);
    }
    \end{pgfonlayer}
    \begin{pgfonlayer}{background}
      \tikzstyle{label}=[font=\scriptsize\sffamily, overlay, fill=white, draw=none,
         inner sep=0pt, outer sep=0pt, text centered,
         anchor=north]
      \foreach \i/\l/\u in {a/1/4, c/1/2, d/3/3, b/5/5, e/1/1} 
      {
        \node[label] at (\i.south) (int\i) {$[\mathsf{\l,\!\textbf{\u}}]$}; 
      }
      \foreach \i/\l/\u in {r/1/6} 
      {
        \node[label, anchor=south] at (\i.north) (int\i) {$[\mathsf{\l,\!\textbf{\u}}]$}; 
      }
    \end{pgfonlayer}
  \end{tikzpicture}
}
\hfill
\subfloat[Interval Propagation]{
  \begin{tikzpicture}[baseline=-4ex, xscale=\xsc, yscale=\ysc, tr/.style={font=\scriptsize, circle,
      draw=MPIIdarkblue, minimum size=4.5mm, inner sep=0pt}, >=stealth]
    \foreach \x/\y/\l in {1/3/r, 0/1/c, 2/1/d, 0/2/a, 2/2/b, 1/0/e} 
    { 
      \node[tr] (\l) at (\x,\y) {$\l$}; 
    }

    \begin{pgfonlayer}{background}
    \foreach \s/\t in {r/a, a/c, a/d, c/e, r/b, d/e, b/c, b/d} 
    {
      \draw[->] (\s) -- (\t);
    }
    \end{pgfonlayer}
    \begin{pgfonlayer}{background}
      \tikzstyle{label}=[font=\scriptsize\sffamily, overlay, fill=white, draw=none,
         rounded corners, inner sep=.75pt, outer sep=0pt, text centered, anchor=north]
      \foreach \i/\l in {d/{\mathsf{[1,\!1][3,\!\textbf{3}]}}, e/{\mathsf{[1,\!\textbf{1}]}},  a/{\mathsf{[1,\!\textbf{4}]}},
        c/{\mathsf{[1,\!\textbf{4}]}}, b/{\mathsf{[1,\!3][5,\!\textbf{5}]}}, c/{\mathsf{[1,\!\textbf{2}]}}} 
      {
        \node[label] at (\i.south) (int\i) {$\l$}; 
      }
      \node[label, anchor=south] at (r.north) (intr) {$\mathsf{[1,\!\textbf{6}]}$}; 
    \end{pgfonlayer}
  \end{tikzpicture}
}
  \caption{Post-Order Interval Assignment}
  \label{fig:po:labeling}
\end{figure*}

\smallskip 
\noindent\textbf{Example.} Consider the augmented example graph depicted in Figure
\ref{fig:po:labeling}a with the virtual root node $r$. In this example, the children of a node are
traversed in lexicographical order, leading to the spanning tree induced by the edges shown in bold in
Figure \ref{fig:po:labeling}b.  The first node to be visited is node $e$, which is assigned
post-order number 1. Node $a$ is visited as soon as its children $\{c,d\}$ have been visited. The
last visited node is the root $r$.

\ssparhd{Tree Indexing} The enabling feature, which makes post-order labeling a common ingredient in
reachability indices, is the resulting \emph{identifier locality}: For every (complete) subtree of
$T$, the ordered identifiers of the included nodes form a contiguous sequence of integers. The
vertex set of any such subtree can thus be compactly expressed as an integer interval. Let $T_v =
(V_{T_v},E_{T_v})$ denote the subtree of $T$ rooted at node $v$. We have
\begin{align}
  \label{eq:treeinterval}
  \bigl\{ \pi(w) \ \big\vert\ w \in V_{T_v} \bigr\} &= \left[ \min_{w \in V_{T_v}} \pi(w), \ \max_{w
      \in V_{T_v}} \pi(w) \right] \\
     &= \left[ \min_{w\in V_{T_v}} \pi(w), \ \pi(v) \right]. \nonumber
\end{align}
Above interval is called \textit{tree interval} of $v$ and will be denoted by $I_T(v)$ in the
remainder of the text. 

\smallskip\noindent\textbf{Example (cont'd).} The subtree rooted at node $a$ in Figure
\ref{fig:po:labeling}b contains the nodes $\{a,c,d,e\}$ with the set of identifiers
$\{4,2,3,1\}$. Thus, the nodes reachable from $a$ in $T$ are represented by the tree interval
$[1,4]$. The final assignment of tree intervals to the nodes is shown in Figure
\ref{fig:po:labeling}c.\finex
\\[2.5pt]
The complete reachability information of the spanning tree $T$ is encoded in the collection of tree
intervals. For a pair of nodes $u,v\in V$, there exists a path from $u$ to $v$ in $T$ iff the
post-order number of the target is contained in the tree interval of the source, that is,
\begin{equation}
  \label{eq:16}
  u \sim_T v \quad\Longleftrightarrow\quad \pi(v) \in I_T(u).
\end{equation}
This reachability index for trees allows for $\mathcal{O}(1)$ query processing at a space
consumption of $\mathcal{O}(n)$.

\ssparhd{Extension to DAGs} While above technique can be used to easily answer reachability queries
on trees, the case of general DAGs is much more challenging. The reason is that, in general, the
reachable set $\rr(v)$ of a vertex $v$ in the DAG is only partly represented by the interval
$I_T(v)$, as the tree interval only accounts for reachability relationships that are preserved in
$T$. Vertices that can only be reached from a node $v$ by traversing one or more non-tree edges have
to be handled seperately: instead of merely storing the tree intervals $I_T(v)$, every node $v$ is
now assigned a \emph{set of intervals}, denoted by $\mathcal{I}(v)$. The purpose of this so-called
\emph{reachable interval set} is to capture the complete reachability information of a node. The
sets $\mathcal{I}(v), v\in V$ are initialized to contain only the tree interval $I_T(v)$. Then, the
vertices are visited in reverse topological order.  For the current vertex $v$ and every incoming
edge $(u,v) \in E$, the reachable interval set $\mathcal{I}(v)$ is merged into the set
$\mathcal{I}(u)$. The merge operation on the intervals resolves all cases of interval subsumption
and extension exhaustively, eventually ensuring interval disjointness. Due to the fact that the
vertices are visited in reverse topological order, it is ensured that for every non-tree edge $(s,t)
\in E \setminus E_T$, the reachability intervals in $\mathcal{I}(t)$ will be propagated and merged
into the reachable interval sets of $s$ \emph{and all its predecessors}. As a result, all
reachability relationships are covered by the resulting intervals.

\smallskip\noindent\textbf{Example (cont'd).} Figure \ref{fig:po:labeling}c depicts the assignment
of tree intervals to the nodes. As described above, in order to compute the reachable interval sets,
the nodes are visited in ascending order of the post-order values (or, equivalently, in reverse
topological order), thus starting at node $e$. The tree interval $I_T(e) = [1,1]$ is merged into the
set of node $c$, leaving $\mathcal{I}(c) = \{[1,2]\}$ unchanged due to interval subsumption. Next,
$I_T(e)$ is merged at node $d$, resulting in the reachable interval set $\mathcal{I}(d) =
\{[1,1],[3,3]\}$. The reverse topological order in which the vertices are visited ensures that the
interval $[1,1]$ is further propagated to the nodes $b,a$, and $r$.\finex

\smallskip\noindent\textbf{Query Processing.} Using the reachable interval sets $\mathcal{I}(v)$,
queries on DAGs can be answered by checking whether the post-order number of the target is contained
in \emph{one of the intervals} associated with the source:
\begin{equation}
  \label{eq:17}
  u \sim v \quad\Longleftrightarrow\quad \exists \bigl([\alpha,\beta] \in \mathcal{I}(u)\bigr) 
  : \alpha \leq \pi(v) \leq \beta. 
\end{equation}

\smallskip
\noindent\textbf{Example.} Consider again the graph depicted in
Figure \ref{fig:po:labeling}d. The reachable vertex set of node $d$ is given by $\mathcal{I}(d) =\{
[1,1],[3,3] \}$.  This set provides all the necessary information in order to answer reachability
queries involving the source node $d$.

By ordering the intervals contained in a set, reachability queries can now be answered efficiently
in $\mathcal{O}(\log n)$ time on DAGs. The resulting index (collection of reachable interval sets)
can be regarded as a materialization of the transitive closure of the graph, rendering this approach
potentially infeasible for large graphs, both in terms of space consumption as well as computational
complexity.


\section{Approximate Intervals}
\label{sec:approx}
For massive problem instances, indexing approaches that materialize the transitive closure (or
compute a compressed variant without an a priori size restriction), suffer from limited
applicability.  For this reason, recent work on reachability query processing over massive graphs
includes a shift towards guided online search procedures. In this setting, every node is assigned a
concise label which -- in contrast to the interval sets described in Section~\ref{sec:interval} --
is restricted by a predefined size constraint. These labels in general do not allow answering the
query after inspection of just the source node, yet can be used to prune portions of the graph in an
online search.

As a basic example, consider a reachability index that labels every node $v\in V$ with its
topological order number $\tau(v)$. While this simple variant of node labeling is obviously not
sufficient to answer a reachability query by means of a single-lookup, a graph search procedure can
greatly benefit from the node labels: For a given query $(s,t)$, the online search rooted at $s$ can
terminate the expansion of a branch of the graph whenever for the currently considered node $v$ it
holds
\begin{equation}
  \label{eq:po-for-pruning}
  \tau(v) \geq \tau(t). 
\end{equation}
This follows from the properties of a topological ordering.

\noindent The recently proposed GRAIL reachability index~\cite{Yildirim2010,Yildirim2011} further
extends this idea by labeling the vertices with approximate intervals:

Suppose that for every node $v$ we replace the set
$\mathcal{I}(v)$ by a single interval
\begin{equation}
  \label{eq:grail-interval}
  I'(v) := \left[ \min_{w \in \mathcal{R}(v)} \pi(w), \max_{w \in \mathcal{R}(v)} \pi(w)\right],
\end{equation}
spanning from the lowest to the highest reachable id. \comment{For node $d$, this interval is given by
$[1,3]$.} This interval is approximate in the sense that all reachable ids are covered whereas false
positive entries are possible:

\medskip
\noindent
\begin{tikzpicture}
  \node[fill=black!15, text justified, text width=\columnwidth-10pt, inner sep=5pt] (text) at (0,0) 
  {\vspace*{-8pt}
\begin{ssdef}[False Positive]
  Let $v\in V$ denote a node with the approximate interval $I'(v) = [\alpha,\beta]$. A vertex $w \in
  V$ is called \emph{false positive} with respect to $I'(v)$ if
 \begin{equation}
 \label{eq:4}
   \alpha \leq \pi(w) \leq \beta \quad\text{and}\quad v \not\sim w.
 \end{equation}
\end{ssdef}
};
\end{tikzpicture}
\medskip

\comment{In above example, node $c$ is a false positive with respect to $I'(d)$ as $\pi(c) \in
  I'(d)$ whereas no path exists from $d$ to $c$.} Obviously, the single interval $I'(v)$ is not
sufficient to establish a definite answer to a reachability query of the form $(G,v,w)$. However,
all queries involving a target id $\pi(w)$ that lies outside the interval, i.\,e.
\begin{equation}
  \label{eq:19}
  \pi(w) <  \alpha \quad\textrm{or}\quad \pi(w) > \beta,
\end{equation}
can be answered instantly with a negative answer, similar to the basic approach based on
Equation~(\ref{eq:po-for-pruning}). In the opposite case, that is,
\begin{equation}
  \label{eq:20}
  \alpha \leq \pi(w) \leq \beta,
\end{equation}
no definite answer to the reachability query can be given and the online search procedure continues
with an expansion of the child vertices, terminating as soon as the target node is encountered or
all branches have been expanded or pruned, respectively.

In practical applications the GRAIL index assigns a number of $k \geq 1$ such approximate intervals
to every vertex, each based on a different (random) spanning tree of the graph. The intuition behind
this labeling is that an ensemble of independently generated intervals improves the effectiveness of
the node labels since each additional interval potentially reduces the remaining false positive entries. \\
The advantage of this indexing approach over a materialization of the transitive closure is obvious:
the size of the resulting labels can be determined a priori by an appropriate selection of the
number ($k$) of intervals assigned to each node. In addition, the node labels are easily computed by
means of $k$ DFS traversals of the graph.

Empirically, GRAIL has been shown to greatly improve the query processing time over online DFS
search in many cases. However, especially in the case of positive queries, a large portion of the
graph still has to be expanded. While extensions have been proposed to GRAIL to improve performance
on positive queries~\cite{Yildirim2011}, the processing time in these cases remains high.
Furthermore, while an increase of the number of intervals assigned to the nodes potentially reduces
false positive elements, no guarantee can be made due to the heuristic nature of the underlying
algorithm. As a result, in many cases superfluous intervals are stored, in some cases
\emph{negatively} impacting query processing time.


\section{The Ferrari Reachability Index}
\label{sec:ferrari}
In this section, we present the \ferrari reachability index which enables fast query processing
performance over massive graphs by a more involved node labeling approach. The main goal of our
index is the assignment of a \emph{mixture of exact and approximate} reachability intervals to the
vertices with the goal of minimizing the expected query processing time, given \emph{a
  user-specified size constraint on the index}. Contrasting previously proposed approaches, we show
both theoretically and empirically that the interval assignment of the \ferrari index utilizes the
available space for maximum effectiveness of the node labels.

Similar to previously proposed index
structures~\cite{Agrawal1989,vanSchaik2011,Yildirim2010,Yildirim2011}, we use intervals to encode
reachability relationships of the vertices. However, in contrast to existing approaches, \ferrari
can be regarded as an \emph{adaptive transitive closure compression} algorithm.
More precisely, \ferrari uses \emph{selective interval set compression}, where a subset of adjacent
intervals in an interval set is merged into a smaller number of approximate intervals. The resulting
node label then retains a high pruning effectiveness under a given size-restriction. \\
Before we delve into the details of our algorithms and the according query processing
procedure, we first introduce the basic concepts that facilitate our interval assignment approach.

The \ferrari index distinguishes between two types of intervals: approximate (similar to the
intervals in Section \ref{sec:approx}) and exact (as in Section \ref{sec:interval}), depending on
whether they contain false positive elements or not. 

Let $I$ denote an interval. To easily distinguish between interval types, we introduce an indicator
variable $\eta_I$ such that
\begin{equation}
  \label{eq:approx-indicator}
  \eta_I := \begin{cases} 0 & \text{if } I \text{ approximate}, \\ 1 & \text{if } 
    I \text{ exact}. \end{cases}
\end{equation}

As outlined above, a main characteristic of \ferrari is the assignment of size-restricted interval
sets comprising approximate and exact intervals as node labels.  Before we introduce the algorithmic
steps that facilitate the index construction, it is important to explain how reachability queries
can be answered using the proposed interval sets. Let $(G,s,t)$ denote a reachability query and
$\mathcal{I}(s) = \{I_1,I_2,\ldots,I_N\}$ the set of intervals associated with node $s$. In order to
determine whether $t$ is reachable from node $s$, we have to check whether the post-order identifier
$\pi(t)$ of $t$ is included in one of the intervals in the set $\mathcal{I}(s)$. If $\pi(t)$ lies
outside of all intervals $I_1,\ldots,I_N$, the query terminates with a negative answer. If however
it holds that $\pi(t) \in I_i$ for one $I_i \in \mathcal{I}(s)$, we have to distinguish two cases:
(i) if $I_i$ is exact then $s$ is guaranteed to reach node $t$ and (ii) if $I_i$ is
approximate, the neighbors of node $s$ have to be queried recursively until a definite answer can be
given. Obviously, recursive expansions are costly and it is thus desirable to
minimize the number of cases that require lookups beyond the source node.\\
To formally introduce the according optimization problem, we define the notion of \emph{interval
  covers}:

\medskip
\noindent
\begin{tikzpicture}
  \node[fill=black!15, text justified, text width=\columnwidth-10pt, inner sep=5pt] (text) at (0,0) 
  {\vspace*{-8pt}
\begin{ssdef}[$k$-Interval Cover]
  Let $k\geq 1$ denote integer and $\mathcal{I} =
  \bigl\{[\alpha_1,\beta_1],\ldots,[\alpha_N,\beta_N]\bigr\}$ a set of intervals. A
  set $\mathcal{C} = \bigl\{[\alpha'_1,\beta'_1],\ldots,[\alpha'_l,\beta'_l]\bigr\}$ is called
  \emph{$k$-interval cover} of $\mathcal{I}$, written as $\mathcal{C} \sqsupseteq_k \mathcal{I}$, if
  $\mathcal{C}$ covers all elements from $\mathcal{I}$ using no more than $k$ intervals, i.\,e.
  \begin{align}
    &\bigcup_{i=1}^N \bigl\{\, j \ \vert\ \alpha_i \leq j \leq \beta_i \bigr\} \subseteq  
    \bigcup_{i=1}^l \bigl\{\, j' \ \vert\ \alpha'_i \leq j' \leq \beta'_i \bigr\} \\[1mm]
    \textrm{with}\quad &l \leq k.
  \end{align}
\end{ssdef}};
\end{tikzpicture}
\medskip

Note that an interval cover of a set of intervals is easily obtained by merging an arbitrary number
of adjacent intervals in the input set. Next, we address the problem of choosing an $k$-interval
cover that maximizes the pruning effectiveness.

\medskip
\noindent
\begin{tikzpicture}
  \node[fill=black!15, text justified, text width=\columnwidth-10pt, inner sep=5pt] (text) at (0,0) 
  {\vspace*{-8pt}
\begin{ssdef}[Optimal $k$-Interval Cover]  Let $k \geq 1$ denote an integer and
  $\mathcal{I} = \{I_1,I_2,\ldots,I_N\}$ an interval set of size $N$.  We define the
  \emph{optimal $k$-interval cover} of $\mathcal{I}$ by
  \begin{equation}
    \label{eq:opt-k-cover}
    \mathcal{I}^{\ast}_k := \argmin_{\mathcal{C} \ :\ \mathcal{I} \sqsubseteq_k \mathcal{C}} \ 
    \sum_{I \in \,\mathcal{C}} (1-\eta_I)\;\vert I\vert,
  \end{equation}
  that is, the cover of $\mathcal{I}$ with no more than $k$ intervals and the minimum number of
  elements in approximate intervals.
\end{ssdef}
};
\end{tikzpicture}
\medskip

Note that by replacing the set of exact reachability intervals $\mathcal{I}(v)$ by its optimal
$k$-interval cover $\mathcal{I}^{\ast}_k(v)$ -- which is then used as the node label in our index --
we retain maximal effectiveness for terminating a query. The reason is that the number
of cases that require recursive querying directly corresponds to the number of elements contained in
approximate intervals.

\subsection{Computing the Optimal Interval Cover}
While the special cases $k=N$ (optimal $k$-interval cover of $\mathcal{I}$ is the set $\mathcal{I}$
itself) and $k=1$ (optimal solution corresponds to the single approximate interval assigned by
GRAIL, see Equation~\ref{eq:grail-interval}) are easily solved, we next introduce an algorithm that
solves the problem for general values of $k$:

\smallskip As hinted above, an interval cover can be computed by selectively merging adjacent
intervals from the original assignment made to the node $v$. In order to derive an algorithm for
computing $\ii^{\ast}_k(v)$, we first transform the interval set at the node $v$ into its dual
representation where the \emph{gaps between intervals} are specified. As usual, let $\mathcal{I} =
\{I_1,I_2,\ldots,I_N\}$ with $I_i = [\alpha_i,\beta_i]$.

The set $\Gamma := \{\gamma_1,\gamma_2,\ldots, \gamma_{N-1}\}, \,\gamma_i =
[\beta_i+1,\alpha_{i+1}-1]$ denotes the \emph{gaps} between the intervals contained in $\ii$:

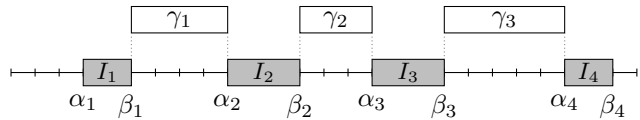
\vspace{3mm}
\begin{center}
  \begin{tikzpicture}[xscale=.32, >=stealth]
    \foreach \f/\t in {0/3,5/9,12/15,18/23,25/26}{ \draw[-] (\f,0) --
      (\t,0);
      \begin{pgfonlayer}{background}
        \foreach \xtick in {\f,...,\t}{ \draw (\xtick,-1.5pt) --
          (\xtick,1.5pt); }
      \end{pgfonlayer}
    }
    \foreach \i/\a/\b in {1/3/5, 2/9/12, 3/15/18, 4/23/25}{
      \begin{pgfonlayer}{background}
        \draw[black!25, line width=10pt] (\a,0) -- (\b,0); \draw
        (\a,0) -- (\a,-5pt) -- (\b,-5pt) -- (\b,5pt) -- (\a,5pt) --
        (\a,0);
      \end{pgfonlayer}
      \node[anchor=north] at (\a,-5pt) {$\alpha_\i$};
      \node[anchor=north] at (\b,-5pt) {$\beta_\i$};
      \node[font=\small] (ivl) at ($(\a,0)!0.5!(\b,0)$) {$I_{\i}$}; }
    \foreach \i/\b/\a in {1/5/9, 2/12/15, 3/18/23}{
      \begin{scope}[yshift=20pt]
        \draw (\b,-5pt) -- (\b,5pt) -- (\a,5pt) -- (\a,-5pt) --
        (\b,-5pt); \node[fill=none] (label) at ($(\b,0)!0.5!(\a,0)$)
        {$\gamma_{\i}$};
        \draw[black!50, densely dotted] (\a,-5pt) -- (\a,-15pt);
        \draw[black!50, densely dotted] (\b,-5pt) -- (\b,-15pt);
      \end{scope}
    }
  \end{tikzpicture}
\end{center}

Note that the gap set $\Gamma$ together with the boundary elements $\alpha_1, \beta_N$ is an
equivalent representation of $\ii$.  For a subset $G \subseteq \Gamma$ we denote by
$\zeta(G)$ the induced interval set obtained by merging adjacent intervals $I_i, I_{i+1}$ if
for their mutually adjacent gap $\gamma_i$ it holds $\gamma_i \notin G$. As an illustrative
example, for the interval set depicted above we have $\zeta\bigl( \{ \gamma_2\}\bigr) := \bigl\{
[\alpha_1,\beta_2], [\alpha_3,\beta_4]\bigr\}$.

Every induced interval set $\zeta(G)$ actually corresponds to a $\vert G\vert +
1$-interval cover of the original set $\ii$. It is easy to see that the \emph{optimal} $k$-interval
cover can equivalently be specified by a subset of gaps.

In order to compute the optimal $k$-interval cover, we thus transform the problem defined in
Equation~(\ref{eq:opt-k-cover}) into the equivalent problem of selecting the ``best'' $k-1$ gaps from
the original gap set $\Gamma$ (or, equivalently, determining the $\vert\mathcal{I}\vert - k-1$ that
are not included in the solution).  For a potential solution $G \subseteq \Gamma$ of at most $k-1$
gaps to preserve, we can assess its \emph{cost}, measured by the number of elements in the induced
interval cover that are contained in approximate intervals:
\begin{equation}
  \label{eq:cost-gap-set}
  c(G) := \sum_{I \in \zeta(G)}  (1-\eta'_I)\;\vert I\vert,
\end{equation}
where for $I \in \zeta(G)$ it holds
\begin{equation}
  \eta'_{I} := \begin{cases} 1 & \textrm{if } I \in \ii \ \wedge\ \eta_{I} = 1, \\ 
    0 & \textrm{else}.\end{cases}
\end{equation}
Clearly, our goal is to determine the set $\Gamma^{\ast}_{k-1} \subseteq \Gamma$ such that
\begin{equation}
  \label{eq:best-gap-selection1}
 \Gamma^{\ast}_{k-1} := \argmin_{G \subseteq \Gamma,\; \vert G\vert \leq k-1}  c(G).
\end{equation}

We now present a dynamic programming approach to obtain the optimal set of $k-1$ gaps. 
In the following, we denote for a sequence of intervals $\mathcal{I} = (I_1,I_2,\ldots,I_N)$, the
subsequence consisting of the first $j$ intervals by $\mathcal{I}_j := (I_1,I_2,\ldots,I_j)$. Now,
observe that every set of gaps $G\subseteq\Gamma, \vert G\vert \leq k-1$ represents a valid
$k$-interval cover for each of the interval sequences $\mathcal{I}_{\max \{ i\! \ \vert\ \!\gamma_i
  \in G\}},\ldots,\mathcal{I}_N$, yet at different costs (the cost corresponding to each of these
coverings is strictly non-decreasing). 
In order to obtain a optimal substructure formulation, consider the problem of computing the optimal
$k$-interval cover for the interval sequence $\mathcal{I}_j$. The possible interval covers can be
represented by a collection of sets of gaps:
\begin{equation}
\label{eq:gap-sets}
\mathcal{G}_{k-1}(\mathcal{I}_j) := \bigl\{ G \subseteq \{\gamma_1,\gamma_2,\ldots,\gamma_{j-1}\} \
\big\vert\ \vert G\vert \leq k-1 \bigr\} = \mathcal{G}^-_{k-1}(\mathcal{I}_j) \cup \mathcal{G}^+_{k-1}(\mathcal{I}_j)
\end{equation}
with
\begin{align}
  \label{eq:constituting-gap-sets}
  &\mathcal{G}^-_{k-1}(\mathcal{I}_j) := \mathcal{G}_{k-1}(\mathcal{I}_{j-1})  \nonumber \\
  \text{and}\quad&\mathcal{G}^+_{k-1}(\mathcal{I}_j) := \bigl\{ G \cup \{\gamma_{j-1}\}\ \big\vert\ G \in
  \mathcal{G}_{k-2}(\mathcal{I}_{j-1}) \bigr\}
\end{align}
that is, $\mathcal{G}^-_{k-1}(\mathcal{I}_j)$ is the collection of all subsets of
$\{\gamma_1,\ldots,\gamma_{j-1}\}$ comprising not more than $k-1$ elements and
$\mathcal{G}^+_{k-1}(\mathcal{I}_j)$ corresponds to the collection of all sets of gaps including
$\gamma_{j-1}$ and not more than $k-2$ elements from $\{\gamma_1,\gamma_2,\ldots,\gamma_{j-2}\}$.

From Equations (\ref{eq:gap-sets},\ref{eq:constituting-gap-sets}) we can deduce that every
$k$-interval cover of $\mathcal{I}_{j}$ and thus the optimal solution is either
\begin{itemize}
\item a $k$-interval cover of $\mathcal{I}_{j-1}$ or
\item a $k-1$-interval cover of $\mathcal{I}_{j-1}$ combined with the gap $\gamma_{j-1}$ between the
  last two intervals, $I_{j-1}$ and $I_j$.
\end{itemize}

Thus, for the optimal solution $\Gamma^\ast_{k-1}(\mathcal{I}_j)$ we have
\begin{align}
  \label{eq:dp-optsub}
  c\bigl(\Gamma^\ast_{k-1}(\mathcal{I}_j)\bigr) &= 
    \min \left\{ \min_{G \in \mathcal{G}^+_{k-1}(\mathcal{I}_j)} c(G),  \min_{G \in
      \mathcal{G}^-_{k-1}(\mathcal{I}_j)} c(G) \right\} \nonumber \\
    &= \min \left\{
         \min_{G \in \mathcal{G}_{k-2}(\mathcal{I}_{j-1})} c(G), \min_{G \in
           \mathcal{G}_{k-1}(\mathcal{I}_{j-2})} c(G) + \vert \gamma_{j-1}\vert + \vert I_j\vert\}
       \right\} \nonumber \\
    &= \min \Big\{ 
      c(\Gamma^\ast_{k-2}\bigl(\mathcal{I}_{j-2})\bigr),\ 
      c\bigl(\Gamma^\ast_{k-1}(\mathcal{I}_{j-2})\bigr) + \vert\gamma_{j-1}\vert + \vert I_j\vert
    \Big\}.
\end{align}

We can exploit the optimal substructure derived in Equation (\ref{eq:dp-optsub}) for the desired
dynamic programming procedure: For each $\mathcal{I}_i, 1 \leq i \leq N$ we have to compute the
$k'$-interval cover for $k-N+i \leq k' \leq k$, thus obtaining the optimal solution in time
$\oh(kN)$.

In some practical applications the amount of computation can become prohibitive, as one instance of
the problem has to be solved for every node in the graph. Thus, in our implementation, we use a
simple and fast greedy algorithm that, starting from the empty set iteratively adds the gap $\gamma
\in \Gamma$ that leads to the greatest reduction in cost given the current selection $G$, until at
most $k-1$ gaps have been selected, then compute the interval cover from $\zeta(G)$.  While the gain
in speed comes at the cost of a potentially suboptimal cover, our experimental
evaluation demonstrates that this approach works well in practice.\\
In the next section, we explain how above node labeling technique is eventually used as a building
block during the reachability index computation.

\subsection{Index Construction}
At precomputation time, the user specifies a certain budget $B = kn, k \geq 1$ of intervals that can
be assigned to the nodes, thus directly controlling the tradeoff between index size/precomputation
time and pruning effectiveness of the respective nodes labels. The subsequent index construction
procedure can be broken down into the following main stages:

\subsubsection{Tree Cover Construction}
Agrawal et al.~\cite{Agrawal1989} propose an algorithm for computing a tree cover that leads to the
minimum number of exact intervals to store.  This tree can be computed in $\mathcal{O}(mn)$ time,
rendering the approach infeasible for the case of massive graphs. While, in principle, heuristics
could be used that are based on centrality measures or estimates of the sizes of reachable
sets~\cite{Cohen1997,Palmer2002}, we settle for a simpler solution that does not yield a certain
approximation guarantee yet performs well in practice. We argue that a good tree cover should cover
as many reachability relationships as possible in the initial tree intervals (see Equation
\ref{eq:treeinterval}). Therefore, every edge included in the tree should provide a connection
between as many pairs of nodes as possible. To this end, we propose the
following procedure to heuristically construct such a tree cover $T$: \\
Let $\tau: V \rightarrow \{1,2,\ldots,n\}$ denote a topological ordering of the vertices, i.\,e. for
all $(u,v) \in E$ it holds $\tau(u) < \tau(v)$.  Such a topological ordering is easily obtained by
the classical textbook algorithm \cite{Cormen2009} in time $\mathcal{O}(m+n)$. We interpret the
topological order number of a vertex as the number of \emph{potential predecessors} in the graph,
because the number of predecessors of a given node is upper bounded by its position in the
topological ordering. For a vertex $v$ with set of predecessors $\mathcal{N}^-(v)$, we select the
edge from node $p\in \mathcal{N}^-(v)$ with highest topological order number for inclusion in the
tree, that is
\begin{equation}
  \label{eq:select-parent}
  p := \argmax_{u \in \mathcal{N}^-(v)} \ \tau(u).
\end{equation}
The intuition is that node $p$ has the highest number of potential predecessors and thus the
selected edge $(p,v)$ has the potential of providing a connection from a large number of nodes to
$v$, eventually leading to a high number of reachability relationships encoded in the resulting tree
intervals. An overview of the tree cover construction algorithm is depicted in Algorithm
\ref{algo:tc}.

 \begin{algorithm2e}[t]
  \SetKwInput{KwData}{Input} 
  \DontPrintSemicolon 
  \LinesNumbered 
  \KwData{directed acyclic graph $G=(V,E)$}
   \Begin{ 
    $T \leftarrow (V_T,E_T) \leftarrow (V, \emptyset)$ \;
    $\tau \leftarrow \textsc{TopologicalSort}(G)$ \;
    \For{$i= n \ \text{\normalfont\bfseries downto }  1$}{
     $\{ v \} \leftarrow \{ u \in V \ \vert \ \tau(u) = i \}$ \;
     \If{$\mathcal{N}^-(v) \ne \emptyset$}{
       $E_T \leftarrow E_T \cup \left( \argmax_{u \in \mathcal{N}^-(v)} \tau(u), v\right)$ \;
     }
    }
    \textrm{\bfseries return} $T$ \;
   }
   \caption{\textsc{TreeCover}$(G)$}
   \label{algo:tc}
 \end{algorithm2e}

\subsubsection{Interval Set Assignment}
As the next step, given the tree cover $T$, indexing proceeds by assigning the exact tree interval
$I_T(v)$, encoding the reachability relationships within the tree at each node $v$. This
interval assignment can be obtained using a single depth-first traversal of $T$. \\
In order to label every node $v$ with a $k$-interval cover $\mathcal{I}'(v)$ of its true reachable
interval set, we visit the vertices of the graph in reverse topological order, that is, starting
from the leaf with highest topological order, proceeding iteratively backwards to the root node. We
initialize for node $v$ the reachable interval interval set as $\mathcal{I}'(v) := \{ I_T(v)
\}$. For the currently visited node $v$ and every edge $(v,w) \in E$, we merge $\mathcal{I}'(w)$
into $\mathcal{I}'(v)$, such that the resulting set of intervals is closed under subsumption and
extension.\footnote{In our implementation we require non-adjacent intervals in the set, that is, for
  $[\alpha_1,\beta_1], [\alpha_2,\beta_2] \in \ii$ it must hold $\beta_1 < \alpha_2$. When sets of
  approximate and exact intervals are merged, the type of the resulting interval is based on several
  factors. For example, when an exact interval is extended by an approximate interval, the result
  will be one long approximate range.}  

Then, in order to satisfy the size restriction of at most $k$ intervals associated with a node, we
replace $\mathcal{I}'(v)$ by its $k$-interval cover which is then stored as the node label of $v$ in
our index.  The complete procedure is shown in detail in Algorithm \ref{algo:indexconst}. It is easy
to see that the resulting index consisting of the sets of approximate and exact intervals
$\mathcal{I}'(v), v \in V$ comprises \emph{at most} $nk = B$ intervals. The upper bound $\sum_{v\in
  V}\vert\mathcal{I}'(v)\vert \leq B$ is usually not tight, i.\,e. in practice, much less than $B$
intervals are assigned. As an example, in the case of leaf nodes or the root vertex, a single
interval suffices.  The name of the algorithm -- \ferrari-L -- thus reflects the fact that a
\emph{local} size restriction, $\vert\mathcal{I}'(v)\vert \leq k$, is satisfied by every interval
set $\mathcal{I}'(v)$.
\\
Note that even though an optimal algorithm can be used to compute the $k$-interval covers, the
optimality of the local result does in general not extend to the global solution, i.\,e. the full
set of node labels. The reason for this is the fact that adjacent intervals that are merged during
the interval cover computation are propagated to the parent nodes. As a result, at the point during
the execution of the algorithm where the interval set of the parent $p$ has to be covered, the
$k$-interval cover is computed without knowledge of the true (exact) reachability intervals of $p$.
More precisely, the input to the covering algorithm is a combination of approximate (thus previously
merged) and exact intervals. Nevertheless, the resulting node labels prove very effective for early
termination of reachability queries, as our experimental evaluation indicates.

\begin{algorithm2e}[t]
  \SetKwInput{KwData}{Input} 
  \DontPrintSemicolon \LinesNumbered
  \KwData{directed,~acyclic~graph~$G$,~interval~budget~$B = kn$} 
  \KwResult{set of at most $k$ approximate and exact reachability intervals $\mathcal{I}'(v)$
    for every node $v\in V$}
  \Begin{ 
    $T \leftarrow \textsc{TreeCover}(G)$ \;
    $I_T \leftarrow \textsc{AssignTreeIntervals}(T)$ \;
    $k \leftarrow \frac{B}{n}$ \;
    \For(\com*[f]{visit nodes in reverse topological order}){$i=n \ \textrm{\normalfont\bfseries to } 1$}{
      $\{ v\} \leftarrow \{v \in V \ \vert \ \tau(v) = i \}$ \;
      $\mathcal{I}'(v) \leftarrow \{ I_T(v) \}$ \;
      \ForEach{$w \in \mathcal{N}^+(v)$}{
        $\mathcal{I}'(v) \leftarrow \mathcal{I}'(v) \oplus \mathcal{I}'(w)$ \com*{merge interval sets} 
      }
      \smallskip
      \com{replace intervals by $k$-interval cover}
      $\mathcal{I}'(v) \leftarrow \textsc{$k$-IntervalCover}(\mathcal{I}'(v))$ \;
    }
    \textrm{\normalfont\bfseries return } $\bigl\{ \mathcal{I}'(v) \ \vert \ v \in V \bigr\}$ \;
  }
  \caption{\textsc{Ferrari-L}$(G,B)$}
  \label{algo:indexconst}
\end{algorithm2e}

To further improve our reachability index, in the next section we propose a variant of the labeling
algorithm that leads to an even better utilization of the available space.

\subsection{Dynamic Budget Allocation}
As mentioned above, the interval assignment as described in Algorithm~\ref{algo:indexconst} usually
leads to a total number of far less than $B$ intervals stored at the nodes. In order to better
exploit the available space, we extend our algorithm by introducing the concept of \emph{deferred
  interval merging} where we can assign more than $k$ intervals to the nodes on the first visit,
potentially requiring to revisit a node at a later stage of the algorithm. \\
The indexing algorithm for this interval assignment variant works as follows: Similar to \ferrari-L,
nodes are visited in reverse topological order and the interval sets of the neighboring nodes are
merged into the interval set $\mathcal{I}'(v)$ for the current vertex $v$. However, in this new
variant, subsequent to merging the interval sets we compute the interval cover comprising at most
$ck$ intervals, given a constant $c\geq 1$. This way, more intervals can be stored in the node
labels. After the $ck$-interval cover has been computed, the vertex $v$ is added to a min-heap
structure where the nodes are maintained in ascending order of their degree. This procedure
continues until the already assigned interval sets sum up to a size of more than $B$ intervals. In
this case, the algorithm repeatedly pops the minimum element from the heap and restricts its
respective interval set by computing the $k$-interval cover. This deferred interval set restriction
is repeated until the number of assigned intervals again satisfies the size constraint $B$. \\
Abive procedure leads to a much better utilization of the available space and thus a better quality
of the resulting reachability index. The improvement comes at the cost of increased index
computation time, in practice the increase is two-fold in the worst-case, negligible in others. Our
experimental evaluation suggests that a value of $c=4$ provides a reasonable tradeoff between
efficiency of construction and resulting index quality. This second indexing variant is shown in
detail in Algorithm~\ref{algo:indexconst-global}. We refer to the algorithm as the \emph{global}
variant (\ferrari-G) as in this case the size constraint is satisfied over all vertices -- in
contrast to the local size constraint of \ferrari-L.

\begin{algorithm2e}[t]
  \SetKwInput{KwData}{Input} 
  \DontPrintSemicolon 
  \LinesNumbered
  \KwData{directed,~acyclic~graph~$G$,~interval~budget~$B = kn$, constant $c \geq 1$} 
  \KwResult{set of approximate and exact reachability intervals $\mathcal{I}'(v)$
    for every node $v\in V$ s.\,t. the total number of intervals is upper-bounded by $B$}
  \Begin{ 
    $T \leftarrow \textsc{TreeCover}(G)$ \;
    $I_T \leftarrow \textsc{AssignTreeIntervals}(T)$ \;
    $H \leftarrow \textsc{InitializeMinHeap}()$ \;
    \smallskip 
    ${s} \leftarrow 0$ \com*{number of currently assigned intervals} 
    \For(\com*[f]{visit nodes in reverse topological order}){$i=n \ \textrm{\normalfont\bfseries to } 1$}{
      $\{ v\} \leftarrow \{v \in V \ \vert \ \tau(v) = i \}$ \;
      $\mathcal{I}'(v) \leftarrow \{ I_T(v) \}$ \;
      \ForEach{$w \in \mathcal{N}^+(v)$}{
        $\mathcal{I}'(v) \leftarrow \mathcal{I}'(v) \oplus \mathcal{I}'(w)$ \com*{merge interval sets} 
      }
      \smallskip
      \com{replace intervals by $ck$-interval cover}
      $\mathcal{I}'(v) \leftarrow \textsc{$ck$-IntervalCover}(\mathcal{I}'(v))$ \;
      ${s} \leftarrow {s} + \vert \mathcal{I}'(v)\vert$ \;
      \If{$\vert \mathcal{I}'(v)\vert > k$}{
        $\textsc{Heap-Push}\bigl(H,v,\vert\mathcal{N}^+(v)\vert\bigr)$ \;
      }
      \While{$s > B$}{
        $w \leftarrow \textsc{Heap-Pop}(H)$ \;
        $\mathcal{I}'(w) \leftarrow \textsc{$k$-IntervalCover}(\mathcal{I}'(w))$ \;
        ${s} \leftarrow {s} - \vert\mathcal{I}'(w)\vert + k$ \;
      }
    }
    \textrm{\normalfont\bfseries return } $\bigl\{ \mathcal{I}'(v) \ \vert \ v \in V \bigr\}$ \;
  }
  \caption{\textsc{Ferrari-G}$(G,B)$}
  \label{algo:indexconst-global}
\end{algorithm2e}

In the next section, we provide more details about our query answering algorithm and additional
heuristics that further speed up query processing over the \ferrari index.

\section{Query Processing and Additional Heuristics}
\label{sec:qp}
The basic query processing over \ferrari's reachability intervals is straightforward and resembles
the basic approach of Agrawal et al.~\cite{Agrawal1989}: For every node $v$, the intervals in the
set $\mathcal{I}'(v)$ are maintained in sorted order. Then, given a reachability query $(G,s,t)$, it
can be determined in $\mathcal{O}(\log \vert\mathcal{I}'(v)\vert)$ time whether the target id
$\pi(t)$ is contained in one of the intervals of the source. The query returns a negative answer ($s
\not\sim t$) if the target id lies outside all of the intervals and a positive answer if it is
contained in one of the exact intervals. Finally, if $\pi(t)$ falls into one of the approximate
intervals, the
neighbors of $s$ are expanded recursively using a DFS search algorithm. \\
Next, we introduce some heuristics that can further speed up query processing.  \vspace*{-2pt}
\subsection{Seed Based Pruning}
It is evident that, in the case of recursive querying, the performance of the algorithm depends on
the number of vertices that have to be expanded during the online search. Nodes with a very high
outdegree are especially costly as they might lead to a large number of recursive queries. In
practice, such high degree nodes are to be expected due to the fact that (i) most of the real-world
graphs in our target applications will follow a power-law degree distribution and (ii) the
condensation graph obtained from the input graph produces high-degree nodes in many cases because
the large strongly connected components usually exhibit a large number of outgoing edges.\\
To overcome this problem, we propose to determine a set of \emph{seed vertices} $S \subseteq V$ and
assign an additional label to every node $v$ in the graph, indicating for every $\sigma \in S$
whether $G$ contains a forward (backward) directed path from $v$ to $s$.\\
This labeling scheme works as follows: Every node will be associated with two sets, $S^-(v)$ and
$S^+(v)$, such that $S^-(v) := \bigl\{ \sigma \in S \ \vert \ \sigma \sim v \bigr\},$ and $S^+(v) :=
\bigl\{ \sigma \in S \ \vert \ v \sim \sigma \bigr\}.$
Next, we describe the procedure for assigning the sets $S^+$: \\
For every node $v$, we initialize $S^+(v) = \{v\}$ if $v\in S$ and $S^+(v) =\emptyset$ otherwise. We
then maintain a FIFO-queue of all vertices, initialized to contain all leaves of the graph. At each step
of the algorithm, the first vertex $v$ is removed from the queue. Then, for every predecessor $u,
(u,v) \in E$ we set $S^+(u) \leftarrow S^+(u) \cup S^+(v)$. If all successors of $u$ have been
processed, $u$ itself is added to the end. The algorithm continues until all nodes of
the graph have been labeled. It is easy to see that above procedure can efficiently be
implemented. The approach
for assignment of the sets $S^-$ is similar (starting from the root nodes).\\
Once assigned, the sets can be used by the query processing algorithm in the following way: For a
query $(G,s,t)$,
\begin{enumerate}
\item if $S^+(s) \cap S^-(t) \ne \emptyset$, then $s \sim t$.
\item if there exists a seed node $\sigma$ s.\,t. $\sigma \in S^-(s)$ and $\sigma \notin S^-(t)$, that is,
  the seed $\sigma$ can reach $s$ but not $t$, the query can be terminated with a negative answer
  ($s \not\sim t$).
\end{enumerate}

In our implementation we choose to elect the $s$ nodes with maximum degree as seed nodes (requiring
a minimum degree of 1). The choice of $s$ can be specified prior to index construction, in our
experiments we set $s = 32$.

\subsection{Pruning Based on Topological Properties}
We enhance the \ferrari index with two additional powerful criteria that allow additional pruning of
certain queries. First, we adopt the effective topological level filter that was proposed by
Y{\i}ld{\i}r{\i}m et al. for the GRAIL index (see~\cite{Yildirim2011} for details). Second, we
maintain the topological order $\tau(v)$ of each vertex $v$ for pruning as shown in
Equation~(\ref{eq:po-for-pruning}). 

Before we proceed to the experimental evaluation of our index, we first give an overview over
previously proposed reachability indexing approaches.


\section{Related Work}
\label{sec:related-work}
Due to the crucial role played by reachability queries in innumerable applications, indices to speed
them up have been subject of active research.  Instead of exhaustively surveying previous results,
we briefly describe some of the key proposals here. For a detailed survey, we direct the reader
to~\cite{Yu2010}.  In this section, we distinguish between reachability query processing techniques
that are able to answer queries using only the label information on nodes specified in the query,
and those which use the index to speed up guided online search over the graph.

Before we proceed, it is worth noting that there are two recent proposals that aim to speed up the
reachability queries from a different direction compared to the standard graph indexing
approaches. First, in~\cite{Jin2012}, authors propose a novel way to compact the graph before
applying any reachability index. Naturally, this technique can be used in conjunction with \ferrari,
hence we consider it orthogonal to the focus of the present paper. The other proposal is to compress
the transitive closure through a carefully optimized word-aligned bitmap encoding of the
intervals~\cite{vanSchaik2011}. The resulting encoding, called PWAH-8, is shown to make the interval
labeling technique of Nuutila~\cite{Nuutila1996} scale to larger graph datasets. In our experiments,
we compare our performance with both Nuutila's Intervals assignment technique as well as the PWAH-8
variant.

\subsection{Direct Reachability Indices}
\nocite{Agrawal1989, Jagadish1990, Chen2008, Wang2006, Jin2008, Jin2011} Indices in this category
answer a reachability query $(G,s,t)$ using just the labels assigned to $s$ and $t$.
Apart from the classical algorithm~\cite{Agrawal1989} described in Section~\ref{sec:prelim}, another
approach based on tree covering~\cite{Wang2006} focused on sparse graphs (targeting near-tree
structures in XML databases), labeling each node with its tree interval and computing the transitive
closure of non-tree edges separately.

Apart from trees to cover the graph being indexed, alternative simple structures such as chains and
paths have also been used. In a chain covering of the graph, a node $u$ can reach $v$ if both belong
to the same chain and $u$ precedes $v$. In~\cite{Jagadish1990} an optimal way to cover the graph
with chains in $\oh(n^3)$ was proposed, later reduced to $\oh(n^2 + dn \sqrt{d})$, where $d$ denotes
the diameter of the graph \cite{Chen2008}. Although chain covers typically generate smaller index
sizes than the interval labeling and can answer queries efficiently, they are very expensive to
build for large graphs.  The PathTree index proposed recently~\cite{Jin2011} combines tree covering
and path covering effectively to build an index that allows for extremely fast reachability query
processing.  Unfortunately, the index size can be extremely large, consuming upto $\oh(np)$ space,
where $p$ denotes the number of paths in the decomposition.

\nocite{Cohen2002, Schenkel2004} Instead of indexing using covering structures, Cohen et
al.~\cite{Cohen2002} introduced 2-Hop labeling which, at each node $u$, maintains a subset of the
node's ancestors and descendants.  Using this, reachability queries between $s$ and $t$ can be
answered by intersecting the descendant set of $s$ with the ancestors of $t$.  This technique was
particularly attractive for query processing within a database system since it can be implemented
efficiently using SQL-statements performing set intersections~\cite{Schenkel2004}.  The main hurdle
in using it for large graphs turns out to be its construction -- optimally selecting the subsets to
label nodes with is an NP-hard problem, and no bounds on the index size can be specified. HOPI
indexing~\cite{Schenkel2004} tried to overcome these issues by clever engineering, using a
divide-and-conquer approach for computing the covering. 3-Hop labeling~\cite{Jin2009} combines the
idea of chain-covering with the 2-Hop strategy to reduce the index size.

\subsection{Accelerating Online Search}
\nocite{Trissl2007, Yildirim2010, Yildirim2011}

From the discussion above, it is evident that accurately capturing the entire transitive closure in
a manner that scales to massive size graphs remains a major challenge. Some of the recent approaches
have taken a different path to utilize scalable indices that can be used to speed up traditional
\emph{online search} to answer reachability queries. In GRIPP~\cite{Trissl2007}, the index maintains
only one interval per node on the tree cover of the graph, but some nodes reachable through non-tree
edges are replicated to improve the coverage.

The recently proposed GRAIL index~\cite{Yildirim2010,Yildirim2011} uses $k$ random trees to cover
the condensed graph, generating as many intervals to label each node with. As we already described
in~Section~\ref{sec:approx}, the query processing proceeds by using the labels to quickly determine
non-reachability, otherwise recursively querying the nodes underneath in the DAG, resulting in a
worst-case query processing performance of $\oh(k (m+n))$.  Although GRAIL was shown to be able to
build indices over massive scale graphs quite efficiently, it suffers from the previously discussed
drawbacks.  In our experiments, we compare various aspects of our \ferrari index against GRAIL which
is, until now, the only technique that deals effectively with massive graphs while satisfying a
user-specified size-constraint.


\section{Experimental Evaluation}
\label{sec:expt}
We conducted an extensive set of experiments in order to evaluate the performance of \ferrari in
comparison with the state of the art reachability indexing approaches, selected based on recent
results. In this paper, we present the results of our comparison with: GRAIL~\cite{Yildirim2011},
PathTree~\cite{Jin2011}, Nuutila's Intervals~\cite{vanSchaik2011}, and
\mbox{PWAH-8}~\cite{vanSchaik2011}.  For all the competing methods, we obtained original source code
from the authors, and set the parameters as suggested in the corresponding publications.

\subsection{Setup}
Fortunately, all indexing methods are implemented using C++, making the comparisons fairly accurate
without any platform-specific artifacts. 
All experiments were conducted using a Lenovo ThinkPad W520 notebook computer equipped with 8 Intel
Core i7 CPUs at 2.30 GHz, and 16 gigabyte of main memory. The operating system in use was a 64-bit
installation of Linux Mint 12 using kernel 3.0.0.22.

\subsection{Methodology}
The metrics we compare on are: 
\begin{enumerate}
\item {{\bf Construction time } for each indexing strategy over each dataset. Since the input to all
    considered algorithms is always a DAG, we do not include the time for computing the condensation
    graph into our measurements.} 
    \item {{\bf Query processing time} for executing 100,000 reachability queries. We consider
        \emph{random} and \emph{positive} sets of queries and report numbers for both workloads
        separately.
    }
\item {{\bf Index size} in memory that each index consumes.  It should be noted that although both
    \ferrari and GRAIL take as input a size restriction parameter, the resulting size of the index
    can be quite different. PathTree, (Nuutila's) Intervals and \mbox{PWAH-8} have no parameterized
    size, and depend entirely on the dataset characteristics.}
\end{enumerate}

\subsection{Datasets}
We used the selection of graph datasets (Table~\ref{table:datasets}a) that, over the recent years,
has become the benchmark set for reachability indexing work. These graphs are classified based on
whether they are small (with 10-100s of thousands of nodes and edges) or large (with millions of
nodes and edges), and dense or sparse. Due to lack of space, we refer to the detailed description of
these datasets in~\cite{Yildirim2011} and~\cite{Jin2011} and, for the same reason, report results
only for a salient subset (the full set of results can be found in the appendix of this paper). We
term these datasets as \emph{benchmark datasets} and present results accordingly in
Section~\ref{subsec:eval-bm}.

\begin{table}[tb]
\centering
 \subfloat[Benchmark Datasets]{
   \begin{tabular*}{.97\columnwidth}{@{\extracolsep{\fill}}lcrrc}
 \toprule
 \textbf{Dataset} & \textbf{Type} & \multicolumn{1}{c}{$\vert V\vert$} &  \multicolumn{1}{c}{$\vert E\vert$}  & \textbf{Source} \\
 \midrule
 ArXiV & small, dense & 6,000 & 66,707 & \cite{Jin2009} \\
 GO & small, dense & 6,793 & 13,361 & \cite{Jin2009} \\
 Pubmed & small, dense & 9,000 & 40,028 & \cite{Jin2009} \\
 Human & small, sparse & 38,811 & 39,816 & \cite{Jin2008} \\
 CiteSeer & large & 693,947 & 312,282 & \cite{Yildirim2011} \\
 Cit-Patents & large & 3,774,768 & 16,518,947 & \cite{Yildirim2011} \\
 CiteSeerX & large  & 6,540,401 & 15,011,260 & \cite{Yildirim2011} \\
 GO-Uniprot & large & 6,967,956 & 34,770,235 & \cite{Yildirim2011} \\
 \bottomrule
\end{tabular*}
 }
\vskip 5pt
\vskip 3pt
\subfloat[Web Datasets]{
   \begin{tabular*}{.97\columnwidth}{@{\extracolsep{\fill}}lrrrr} 
  \toprule
  \textbf{Dataset} & \multicolumn{1}{c}{$\vert V\vert$} & \multicolumn{1}{c}{$\vert E\vert$} & \multicolumn{1}{c}{$\vert V_C\vert$} & \multicolumn{1}{c}{$\vert E_C\vert$} \\
  \midrule
  GovWild & 8,027,601 & 26,072,221 & 8,022,880 & 23,652,610 \\
  YAGO2 & 16,375,503 & 32,962,379 & 16,375,503 & 25,908,132 \\
  Twitter & 54,981,152 & 1,963,263,821 & 18,121,168 & 18,359,487  \\
  Web-UK & 133,633,040 & 5,507,679,822 & 22,753,644 & 38,184,039 \\
  \bottomrule
\end{tabular*}
}
\caption{Datasets Used}
\label{table:datasets}
\end{table}

In order to evaluate the performance of the algorithms under real-world settings, where
massive-scale graphs are encountered, we use additional datasets derived from publicly available
sources\footnote{\fontsize{7pt}{7pt}\selectfont We used the version of the files provided at
  \url{http://code.google.com/p/grail/}}. These include RDF data, an online social network, and a
World Wide Web crawl. To the best of our knowledge, these constitute some of the largest graphs used
in evaluating the effectiveness of reachability indices to this date. In the following, we briefly
describe each of them, and summarize the key characteristics of these datasets in
Table~\ref{table:datasets}b.

\begin{itemize}
\item{ {\bf GovWild} is a large RDF data collection consisting of about 26 million triples
    representing relations between more than 8 million
    entities.\footnote{\fontsize{7pt}{7pt}\selectfont\url{http://govwild.hpi-web.de/project/govwild-project.html}}
  }
\item{ {\bf Yago2} is another large-scale RDF dataset representing an automatically constructed
    knowledge graph~\cite{Hoffart2012}. The version we used contained close to 33 million edges
    (facts) between 16.3 million nodes
    (entities).\footnote{\fontsize{7pt}{7pt}\selectfont\url{http://www.mpi-inf.mpg.de/yago-naga/yago/}}
  }
\item { The {\bf Twitter} graph \cite{Cha2010} is a representative of a large-scale social
    network. This graph, obtained from a crawl of \texttt{twitter.com}, represents the follower relationship
    between about 50 million
    users.\footnote{{\fontsize{7pt}{7pt}\selectfont\url{http://twitter.mpi-sws.org/}}}}
\item { {\bf Web-UK} is an example of a web graph dataset~\cite{Boldi2008}. This graph
    contains about 133 million nodes (hosts) and 5.5 billion edges
    (hyperlinks).\footnote{\fontsize{7pt}{7pt}\selectfont\url{http://law.di.unimi.it/webdata/uk-union-2006-06-2007-05/}}
  }
\end{itemize}
We present the results of our evaluation over these web-scale graphs in Section~\ref{subsec:eval-web}.

\subsection{Results over Benchmark Graphs}
\label{subsec:eval-bm}
Tables~\ref{tab:exp-benchmark}a-d and the charts in
Figures~\ref{fig:plot-ct-benchmark}--\ref{fig:plot-qt-p-benchmark} summarize the results for the
selected set of benchmark graphs. In the tables, we provide the absolute values -- time in milliseconds
and index size in KBytes, while the figures help to visualize the relative performance of algorithms
over different datasets. In all the tables missing values are marked as ``$-$'' whenever a
dataset could not be indexed by the corresponding strategy -- either due to memory exhaustion or
for taking too long to index (timeout set to $1$M milliseconds). The best performing strategy for
each dataset is shown in bold. For GRAIL we set the number of dimensions as suggested
in~\cite{Yildirim2011}, that is, to 2 for small sparse graphs (Human), 3 for small dense graphs
(ArXiV, GO, PubMed) and to 5 for the remaining large graphs. The input parameter value for \ferrari
was also set correspondingly for a fair comparison.

\subsubsection{Index Construction}
Table~\ref{tab:exp-benchmark}a and Figure~\ref{fig:plot-ct-benchmark} present the construction time
for the various algorithms. The results show that the GRAIL index can be constructed very
efficiently on small graphs, irrespective of the density of the graph. On the other hand, the
performance of PathTree is highly sensitive to the density of the graph as well as the size. While
GRAIL and FERRARI's indexing time increases corresponding to the size of the graphs, PathTree simply
failed to complete building for 3 of the larger graphs -- Cit-Patents and CiteSeerX due to memory
exhaustion, GO-Uniprot due to timeout. 

The transitive closure compression algorithms Interval and PWAH-8 can index quite efficiently even
the large graphs and their index is also surprisingly compact. A remarkable exception to this
is the behavior on the Cit-Patents dataset, which seems to be by far the most difficult graph for
reachability indexing. The Interval index failed to process the graph within the given time
limit. The related PWAH-8 algorithm finished the labeling only after around 12 minutes and ended up
generating the \emph{largest index} in all our experiments (including the indices for the Web
graphs). This is rather surprising, as both algorithms were able to index the larger and denser
GO-Uniprot.

\begin{table}
\centering
\subfloat[Construction Time (ms)]{
\begin {tabular*}{\textwidth}{@{\extracolsep{\fill}}rrrrrrr}%
\toprule \bfseries Dataset&\bfseries Ferrari-L&\bfseries Ferrari-G&\bfseries Grail&\bfseries PathTree&\bfseries Interval&\bfseries PWAH-8\\\midrule %
ArXiV&\pgfutilensuremath {15.84}&\pgfutilensuremath {26.62}&$\mathbf {\pgfmathprintnumber {7.862}}$&\pgfutilensuremath {4{,}537.39}&\pgfutilensuremath {34.54}&\pgfutilensuremath {70.10}\\%
Pubmed&\pgfutilensuremath {14.28}&\pgfutilensuremath {24.54}&$\mathbf {\pgfmathprintnumber {8.205}}$&\pgfutilensuremath {326.54}&\pgfutilensuremath {20.35}&\pgfutilensuremath {44.41}\\%
Human&\pgfutilensuremath {23.36}&\pgfutilensuremath {23.37}&\pgfutilensuremath {15.93}&\pgfutilensuremath {348.48}&$\mathbf {\pgfmathprintnumber[fixed,precision=2,zerofill] {2.7000}}$&\pgfutilensuremath {3.82}\\%
GO&\pgfutilensuremath {6.48}&\pgfutilensuremath {6.91}&$\mathbf {\pgfmathprintnumber {4.83}}$&\pgfutilensuremath {89.83}&\pgfutilensuremath {5.06}&\pgfutilensuremath {8.67}\\%
CiteSeer&\pgfutilensuremath {450.12}&\pgfutilensuremath {459.90}&\pgfutilensuremath {2{,}015.90}&\pgfutilensuremath {26{,}479.70}&$\mathbf {\pgfmathprintnumber[fixed,precision=2,zerofill] {251.0970}}$&\pgfutilensuremath {416.41}\\%
CiteSeerX&\pgfutilensuremath {14{,}110.20}&\pgfutilensuremath {16{,}233.40}&\pgfutilensuremath {20{,}528.40}&--&$\mathbf {\pgfmathprintnumber {5808.7870}}$&\pgfutilensuremath {14{,}444.09}\\%
GO-Uniprot&\pgfutilensuremath {26{,}105.90}&\pgfutilensuremath {29{,}611.90}&\pgfutilensuremath {34{,}518.40}&--&$\mathbf {\pgfmathprintnumber {15213.5470}}$&\pgfutilensuremath {26{,}745.61}\\%
Cit-Patents&$\mathbf {\pgfmathprintnumber[fixed,precision=2,zerofill] {20665.5}}$&\pgfutilensuremath {32{,}366.20}&\pgfutilensuremath {21{,}621.70}&--&--&\pgfutilensuremath {751{,}984.08}\\\bottomrule %
\end {tabular*}%

}
\vspace*{.25cm}
\subfloat[Index Size (Kb)]{
\begin {tabular*}{\textwidth}{@{\extracolsep{\fill}}rrrrrrr}%
\toprule \bfseries Dataset&\bfseries Ferrari-L&\bfseries Ferrari-G&\bfseries Grail&\bfseries PathTree&\bfseries Interval&\bfseries PWAH-8\\\midrule %
ArXiV&$\mathbf {\pgfmathprintnumber {243.85546875}}$&\pgfutilensuremath {275.33}&\pgfutilensuremath {304.69}&\pgfutilensuremath {338.07}&\pgfutilensuremath {1{,}364.99}&\pgfutilensuremath {315.24}\\%
Pubmed&$\mathbf {\pgfmathprintnumber {283.67578125}}$&\pgfutilensuremath {413.06}&\pgfutilensuremath {457.03}&\pgfutilensuremath {419.03}&\pgfutilensuremath {1{,}523.83}&\pgfutilensuremath {358.96}\\%
Human&\pgfutilensuremath {768.88}&\pgfutilensuremath {770.30}&\pgfutilensuremath
{1{,}364.45}&\pgfutilensuremath {458.01}&\pgfutilensuremath {160.56}&$\mathbf {\pgfmathprintnumber {160.22}}$\\%
GO&\pgfutilensuremath {200.37}&\pgfutilensuremath {251.01}&\pgfutilensuremath {344.96}&
\pgfutilensuremath {133.30}&\pgfutilensuremath {180.58}&$\mathbf {\pgfmathprintnumber {81.86}}$\\%
CiteSeer&\pgfutilensuremath {13{,}933.90}&\pgfutilensuremath {13{,}934.29}&\pgfutilensuremath
{56{,}925.34}& 
\pgfutilensuremath {9{,}221.61}&\pgfutilensuremath {7{,}733.94}&$\mathbf {\pgfmathprintnumber {6723.36}}$\\%
CiteSeerX&\pgfutilensuremath {158{,}046.72}&\pgfutilensuremath {242{,}236.08}&\pgfutilensuremath {536{,}517.27}&--&\pgfutilensuremath {430{,}913.36}&$\mathbf {\pgfmathprintnumber[fixed,precision=2,zerofill] {152354.441406}}$\\%
GO-Uniprot&\pgfutilensuremath {429{,}564.04}&\pgfutilensuremath {442{,}301.79}&\pgfutilensuremath {571{,}590.14}&--&\pgfutilensuremath {774{,}081.33}&$\mathbf {\pgfmathprintnumber[fixed,precision=2,zerofill] {249883.804688}}$\\%
Cit-Patents&$\mathbf {\pgfmathprintnumber[fixed,precision=2,zerofill] {151631.730469}}$&\pgfutilensuremath {239{,}609.23}&\pgfutilensuremath {309{,}648.94}&--&--&\pgfutilensuremath {5{,}462{,}135.76}\\\bottomrule %
\end {tabular*}%

}
\vspace*{.25cm}
\subfloat[Query Processing Performance (ms), 100k random queries]{
 \begin {tabular*}{\textwidth}{@{\extracolsep{\fill}}rrrrrrr}%
\toprule \bfseries Dataset&\bfseries Ferrari-L&\bfseries Ferrari-G&\bfseries Grail&\bfseries PathTree&\bfseries Interval&\bfseries PWAH-8\\\midrule %
ArXiV&\pgfutilensuremath {23.69}&\pgfutilensuremath {13.91}&\pgfutilensuremath {100.92}&$\mathbf {\pgfmathprintnumber {3.414}}$&\pgfutilensuremath {4.17}&\pgfutilensuremath {23.22}\\%
Pubmed&\pgfutilensuremath {7.58}&\pgfutilensuremath {4.88}&\pgfutilensuremath {12.27}&$\mathbf {\pgfmathprintnumber {2.758}}$&\pgfutilensuremath {3.16}&\pgfutilensuremath {28.58}\\%
Human&$\mathbf {\pgfmathprintnumber {0.784}}$&$\mathbf {\pgfmathprintnumber {0.777}}$&\pgfutilensuremath {4.98}&\pgfutilensuremath {1.21}&\pgfutilensuremath {1.07}&\pgfutilensuremath {1.06}\\%
GO&\pgfutilensuremath {4.10}&\pgfutilensuremath {2.96}&\pgfutilensuremath {4.83}&$\mathbf {\pgfmathprintnumber {2.038}}$&\pgfutilensuremath {2.47}&\pgfutilensuremath {4.45}\\%
CiteSeer&\pgfutilensuremath {6.13}&\pgfutilensuremath {6.24}&\pgfutilensuremath {8.05}&$\mathbf {\pgfmathprintnumber {5.011}}$&\pgfutilensuremath {8.28}&\pgfutilensuremath {12.39}\\%
CiteSeerX&\pgfutilensuremath {15.88}&\pgfutilensuremath {9.31}&\pgfutilensuremath {41.23}&--&$\mathbf {\pgfmathprintnumber {9.2740}}$&\pgfutilensuremath {21.32}\\%
GO-Uniprot&\pgfutilensuremath {28.30}&\pgfutilensuremath {28.92}&$\mathbf {\pgfmathprintnumber {5.9440}}$&--&\pgfutilensuremath {16.82}&\pgfutilensuremath {48.70}\\%
Cit-Patents&\pgfutilensuremath {778.09}&$\mathbf {\pgfmathprintnumber[fixed,precision=2,zerofill] {502.197}}$&\pgfutilensuremath {578.83}&--&--&\pgfutilensuremath {1{,}514.91}\\\bottomrule %
\end {tabular*}%

}
\vspace*{.25cm}
\subfloat[Query Processing Performance (ms), 100k positive queries]{
 \begin {tabular*}{\textwidth}{@{\extracolsep{\fill}}rrrrrrr}%
\toprule \bfseries Dataset&\bfseries Ferrari-L&\bfseries Ferrari-G&\bfseries Grail&\bfseries PathTree&\bfseries Interval&\bfseries PWAH-8\\\midrule %
ArXiV&\pgfutilensuremath {62.64}&\pgfutilensuremath {37.98}&\pgfutilensuremath {220.31}&$\mathbf {\pgfmathprintnumber {4.937}}$&\pgfutilensuremath {5.95}&\pgfutilensuremath {17.74}\\%
Pubmed&\pgfutilensuremath {31.31}&\pgfutilensuremath {20.28}&\pgfutilensuremath {85.38}&$\mathbf {\pgfmathprintnumber {4.415}}$&\pgfutilensuremath {6.21}&\pgfutilensuremath {43.58}\\%
Human&\pgfutilensuremath {2.08}&\pgfutilensuremath {1.96}&\pgfutilensuremath {14.48}&$\mathbf {\pgfmathprintnumber[fixed,precision=2,zerofill] {1.296}}$&\pgfutilensuremath {1.79}&\pgfutilensuremath {6.07}\\%
GO&\pgfutilensuremath {10.72}&\pgfutilensuremath {4.64}&\pgfutilensuremath {19.59}&$\mathbf {\pgfmathprintnumber {2.044}}$&\pgfutilensuremath {3.26}&\pgfutilensuremath {11.43}\\%
CiteSeer&\pgfutilensuremath {13.37}&\pgfutilensuremath {13.47}&\pgfutilensuremath {85.22}&$\mathbf {\pgfmathprintnumber {6.119}}$&\pgfutilensuremath {15.17}&\pgfutilensuremath {30.60}\\%
CiteSeerX&\pgfutilensuremath {82.76}&\pgfutilensuremath {43.06}&\pgfutilensuremath {700.49}&--&$\mathbf {\pgfmathprintnumber {30.3820}}$&\pgfutilensuremath {69.21}\\%
GO-Uniprot&\pgfutilensuremath {65.00}&\pgfutilensuremath {64.72}&\pgfutilensuremath {131.46}&--&$\mathbf {\pgfmathprintnumber {31.7550}}$&\pgfutilensuremath {54.55}\\%
Cit-Patents&\pgfutilensuremath {4{,}086.21}&\pgfutilensuremath {2{,}667.38}&\pgfutilensuremath {5{,}409.82}&--&--&$\mathbf {\pgfmathprintnumber[fixed,precision=2,zerofill] {1739.2990}}$\\\bottomrule %
\end {tabular*}%

}
\caption{Experimental Evaluation on Benchmark Datasets}
\label{tab:exp-benchmark}
\end{table}

\def\colferraril{MPIIdarkblue}
\def\colferrarig{MPIIlightgreen}
\def\colgrail{CornflowerBlue}
\def\colpt{BrickRed}
\def\colpwahi{Dandelion}
\def\colpwah{Sepia}
\def\bw{5pt}
\def\ph{5.3cm} 

\begin{landscape}
\hspace*{1.5cm}
\begin{figure*}[h]
  \centering
  \begin{tikzpicture}
    \begin{semilogyaxis}[width=1.4\textwidth, height=\ph, ybar, bar width=\bw, ymin=0, 
      symbolic x coords={ArXiV,Pubmed,Human,GO,CiteSeer,CiteSeerX,GO-Uniprot,Cit-Patents},
      xticklabels={\xtl{ArXiV}, \xtl{Pubmed}, \xtl{Human}, \xtl{GO}, \xtl{CiteSeer}, 
        \xtl{CiteSeerX}, \xtl{GO-Uniprot}, \xtl{Cit-Patents}},     
      xtick=data, 
      ylabel={\footnotesize\sffamily Construction Time (ms)}, ytick={10,100,1000,10000,100000,1000000}, ymajorgrids=true, yminorgrids=false, enlarge y limits=.01,
      legend style={at={(.5,1.1)}, fill=white, inner sep=1pt, anchor=south, legend columns=-1,
      font=\fontsize{7pt}{7pt}\selectfont\sf,
     draw=none},
     legend entries={Ferrari-L\hspace*{2em}, Ferrari-G\hspace*{2em}, Grail\hspace*{2em}, PathTree\hspace*{2em},  Interval\hspace*{2em}, PWAH-8}
]
      \addplot[fill=\colferraril!100] table[y=ferrari-L-s32, x=dataset] {data/overall-ct.dat};
      \addplot[fill=\colferrarig!100] table[y=ferrari-G-s32, x=dataset] {data/overall-ct.dat};
      \addplot[fill=\colgrail!100] table[y=grail, x=dataset] {data/overall-ct.dat};
      \addplot[fill=\colpt!100] table[y=pt, x=dataset] {data/overall-ct.dat};
      \addplot[fill=\colpwahi!100] table[y=pwah-interval, x=dataset] {data/overall-ct.dat};
      \addplot[fill=\colpwah!100] table[y=pwah-8, x=dataset] {data/overall-ct.dat};
     \end{semilogyaxis}
   \end{tikzpicture}
  \caption{Index Construction Time}
  \label{fig:plot-ct-benchmark}
\end{figure*}
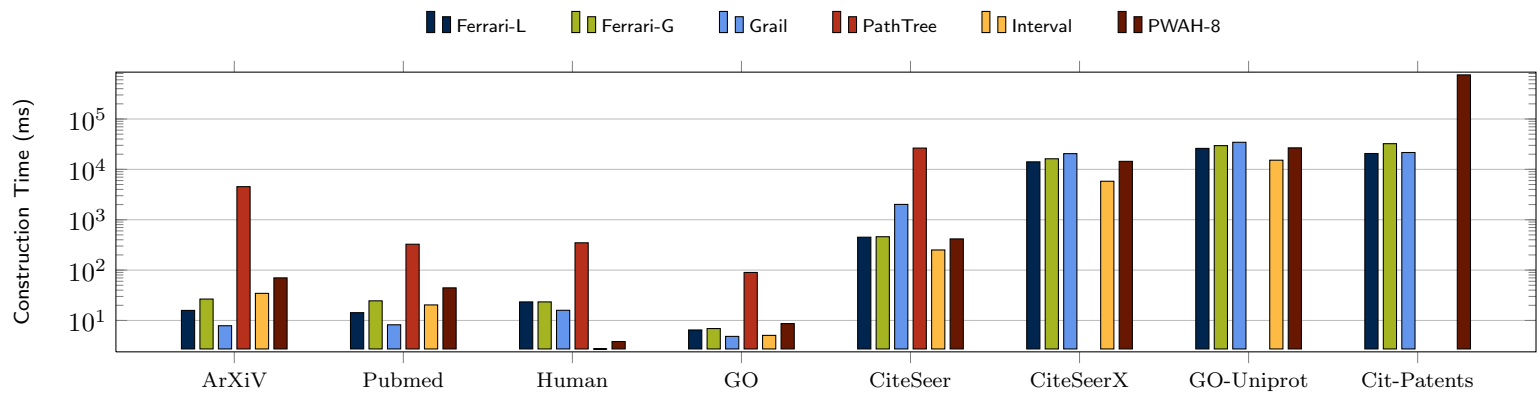
\vspace*{2.5cm}
\begin{figure*}[h]
  \centering
  \begin{tikzpicture}
    \begin{axis}[width=1.4\textwidth, height=\ph, ybar, bar width=\bw, ymin=0, ymax=1.01, 
      symbolic x coords={ArXiV,Pubmed,Human,GO,CiteSeer,CiteSeerX,GO-Uniprot,Cit-Patents},
      xticklabels={\xtl{ArXiV}, \xtl{Pubmed}, \xtl{Human}, \xtl{GO}, \xtl{CiteSeer}, 
        \xtl{CiteSeerX}, \xtl{GO-Uniprot}, \xtl{Cit-Patents}},     
      xtick=data, ylabel={\footnotesize\sffamily Relative Index Size},  xmajorgrids=false, ymajorgrids=true,
]
      \addplot[fill=\colferraril!100] table[y=ferrari-L-s32, x=dataset] {data/overall-size_rel.dat};
      \addplot[fill=\colferrarig!100] table[y=ferrari-G-s32, x=dataset] {data/overall-size_rel.dat};
      \addplot[fill=\colgrail!100] table[y=grail, x=dataset] {data/overall-size_rel.dat};
      \addplot[fill=\colpt!100] table[y=pt, x=dataset] {data/overall-size_rel.dat};
      \addplot[fill=\colpwahi!100] table[y=pwah-interval, x=dataset] {data/overall-size_rel.dat};
      \addplot[fill=\colpwah!100] table[y=pwah-8, x=dataset] {data/overall-size_rel.dat};
     \end{axis}
   \end{tikzpicture}
  \caption{Index Space Consumption}
  \label{fig:plot-size-benchmark}
\end{figure*}
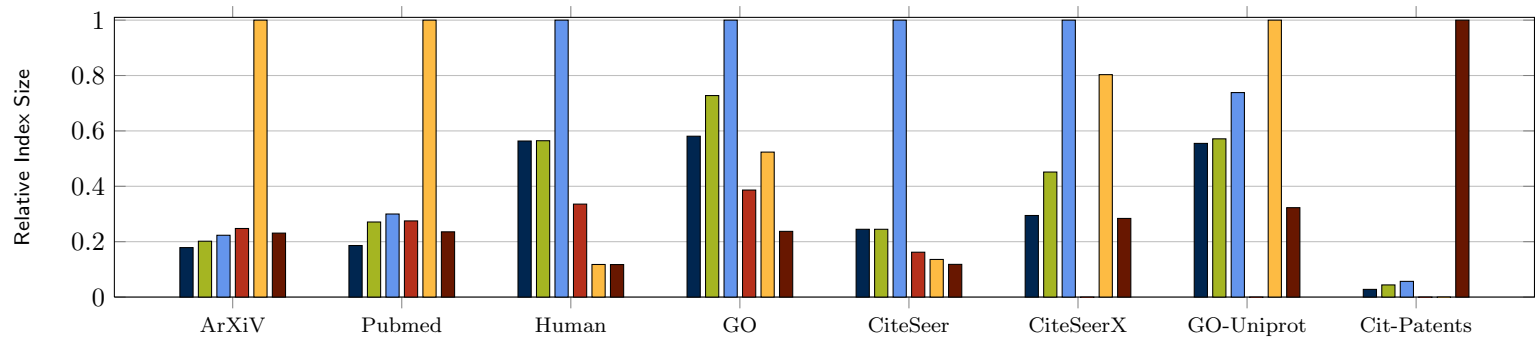
\end{landscape}

\begin{landscape}
\hspace*{1.5cm}
\begin{figure*}
  \centering
  \begin{tikzpicture}
    \begin{axis}[width=1.4\textwidth, height=\ph, ybar, bar width=\bw, ymin=0, ymax=1.01, 
      symbolic x coords={ArXiV,Pubmed,Human,GO,CiteSeer,CiteSeerX,GO-Uniprot,Cit-Patents},
      xticklabels={\xtl{ArXiV}, \xtl{Pubmed}, \xtl{Human}, \xtl{GO}, \xtl{CiteSeer}, 
        \xtl{CiteSeerX}, \xtl{GO-Uniprot}, \xtl{Cit-Patents}},     
      xtick=data, ylabel={\footnotesize\sffamily Relative Time Consumption}, xmajorgrids=false,
      ymajorgrids=true,
      legend style={at={(.5,1.1)}, fill=white, inner sep=1pt, anchor=south, legend columns=-1,
      font=\fontsize{7pt}{7pt}\selectfont\sf,
     draw=none},
     legend entries={Ferrari-L\hspace*{2em}, Ferrari-G\hspace*{2em}, Grail\hspace*{2em}, PathTree\hspace*{2em},  Interval\hspace*{2em}, PWAH-8}
      ]
      \addplot[fill=\colferraril!100] table[y=ferrari-L-s32, x=dataset] {data/overall-qt-r_rel.dat};
      \addplot[fill=\colferrarig!100] table[y=ferrari-G-s32, x=dataset] {data/overall-qt-r_rel.dat};
      \addplot[fill=\colgrail!100] table[y=grail, x=dataset] {data/overall-qt-r_rel.dat};
      \addplot[fill=\colpt!100] table[y=pt, x=dataset] {data/overall-qt-r_rel.dat};
      \addplot[fill=\colpwahi!100] table[y=pwah-interval, x=dataset] {data/overall-qt-r_rel.dat};
      \addplot[fill=\colpwah!100] table[y=pwah-8, x=dataset] {data/overall-qt-r_rel.dat};
     \end{axis}
   \end{tikzpicture}
  \caption{Query Processing Times for 100k Random Queries}
  \label{fig:plot-qt-r-benchmark}
\end{figure*}
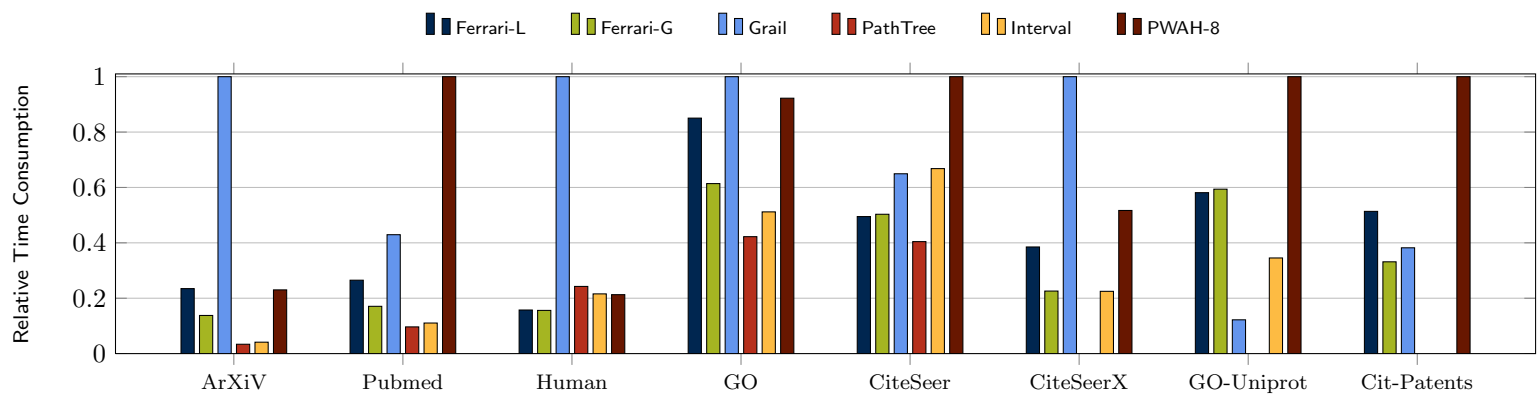
\vspace*{2.5cm}
\begin{figure*}[h]
  \centering
  \begin{tikzpicture}
    \begin{axis}[width=1.4\textwidth, height=\ph, ybar, bar width=\bw, ymin=0, ymax=1.01, 
      symbolic x coords={ArXiV,Pubmed,Human,GO,CiteSeer,CiteSeerX,GO-Uniprot,Cit-Patents},
      xticklabels={\xtl{ArXiV}, \xtl{Pubmed}, \xtl{Human}, \xtl{GO}, \xtl{CiteSeer}, 
        \xtl{CiteSeerX}, \xtl{GO-Uniprot}, \xtl{Cit-Patents}},     
      xtick=data, ylabel={\footnotesize\sffamily Relative Time Consumption},  xmajorgrids=false, ymajorgrids=true,
      ]
      \addplot[fill=\colferraril!100] table[y=ferrari-L-s32, x=dataset] {data/overall-qt-p_rel.dat};
      \addplot[fill=\colferrarig!100] table[y=ferrari-G-s32, x=dataset] {data/overall-qt-p_rel.dat};
      \addplot[fill=\colgrail!100] table[y=grail, x=dataset] {data/overall-qt-p_rel.dat};
      \addplot[fill=\colpt!100] table[y=pt, x=dataset] {data/overall-qt-p_rel.dat};
      \addplot[fill=\colpwahi!100] table[y=pwah-interval, x=dataset] {data/overall-qt-p_rel.dat};
      \addplot[fill=\colpwah!100] table[y=pwah-8, x=dataset] {data/overall-qt-p_rel.dat};
     \end{axis}
   \end{tikzpicture}
  \caption{Query Processing Times for 100k Positive Queries}
  \label{fig:plot-qt-p-benchmark}
\end{figure*}
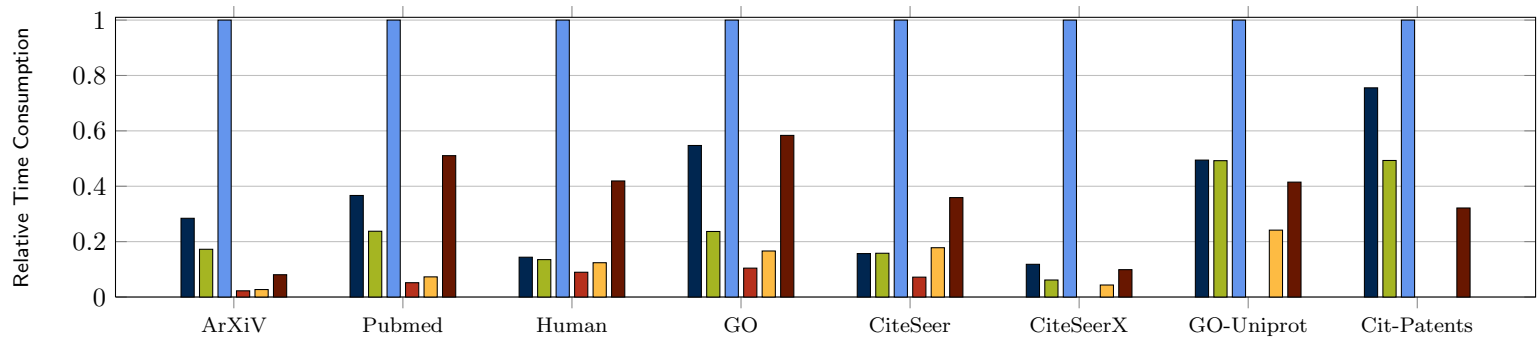
\end{landscape}

When compared to other algorithms, the construction times of \ferrari-L and \ferrari-G are highly
scalable and are not affected much by the variations in the density of the graph. On all graphs,
\ferrari constructs the index quickly, while maintaining a competitive index size. For the
challenging Cit-Patents dataset, it generates the most compact index among all the techniques
considered, and very fast -- amounting to a 23x-36x speedup over PWAH-8. Further, \ferrari
consistently generates smaller indices than GRAIL, and exhibits comparable indexing time. With a few
more clever engineering tricks (e.\,g., including the PWAH-8-style interval encoding), it should be
possible to further reduce the size of \ferrari.

\subsubsection{Query Processing}
Moving on to query processing, we consider random and positive query workloads, with results
depicted in Figures~\ref{fig:plot-qt-r-benchmark} and~\ref{fig:plot-qt-p-benchmark}
(Tables~\ref{tab:exp-benchmark}c and \ref{tab:exp-benchmark}d), respectively. These results help to
highlight the consistency of \ferrari in being able to efficiently process both types of queries
over all varieties of graphs very efficiently. Although, for really small graphs, PathTree is the
fastest as we explained above, it cannot be applied on larger datasets. As graphs get larger, the
Interval indexing turns out to be the fastest. This is not very surprising, since Interval
materializes the exact transitive closure of the graph. \ferrari-G consistently provides competitive
query processing times for both positive as well as random queries over all datasets.  As a
remarkable result of our experimental evaluation consider the CiteSeerX dataset. In this setting,
the Interval index consumes almost twice as much space as the corresponding \ferrari-G index, yet is
only faster by 0.04 milliseconds for random and 12.68 milliseconds for positive queries.

\subsection{Evaluation over Web Datasets}
\label{subsec:eval-web}
As we already pointed out in the introduction, our goal was to develop an index that is both compact
and efficient for use in many analytics tasks when the graphs are of web-scale. For this reason, we
have collected graphs that amount to up to 5 billions of edges before computing the condensation
graph. These graphs are of utmost importance because the resulting DAG exhibits special properties
absent from previously considered benchmark datasets. In this section, we present the results of
this evaluation. Due to its limited scalability, we do not use PathTree index in these
experiments. Also, through initial trial experiments, we found that for GRAIL the suggested
parameter value of $k=5$ does not appear to be the optimal choice, so instead we
report the results with the setting $k=2$, which is also used for \ferrari. The summary of results
is provided in Table~\ref{tab:exp-web}.

\subsubsection{Index Construction}
When we consider the index construction statistics in Tables~\ref{tab:exp-web}a
and~\ref{tab:exp-web}b, it seems that there is no single strategy that is superior across the board.
However, a careful look into these charts further emphasizes the superiority of \ferrari in terms of
its consistent performance. While GRAIL can be constructed fast, its size can be quite large (e.\,g.,
in the case of Twitter).  On the other hand, PWAH-8 can take an order of magnitude more time to
construct than \ferrari as well as GRAIL as we notice for Web-UK. In fact, the Interval index which
is much smaller than \ferrari for Twitter and Yago2, fails to complete within the time allotted for
the Web-UK dataset. In contrast, \ferrari and GRAIL are able to handle any form of graph easily in a
scalable manner.

\begin{table}
 \centering
\subfloat[Construction Time (ms)]{
\begin {tabular*}{\textwidth}{@{\extracolsep{\fill}}rrrrrrr}%
\toprule \bfseries Dataset&\bfseries Ferrari-L&\bfseries Ferrari-G&\bfseries Grail&\bfseries Interval&\bfseries PWAH-8\\\midrule %
YAGO2&\pgfutilensuremath {27{,}713.50}&\pgfutilensuremath {26{,}865.30}&\pgfutilensuremath {17{,}163.00}&$\mathbf {\pgfmathprintnumber {5844.8740}}$&\pgfutilensuremath {9{,}236.71}\\%
GovWild&\pgfutilensuremath {12{,}998.80}&\pgfutilensuremath {18{,}045.30}&$\mathbf {\pgfmathprintnumber {6756.67}}$&\pgfutilensuremath {15{,}060.55}&\pgfutilensuremath {20{,}703.06}\\%
Twitter&\pgfutilensuremath {13{,}065.40}&\pgfutilensuremath {13{,}897.20}&\pgfutilensuremath {9{,}717.39}&\pgfutilensuremath {36{,}480.57}&$\mathbf {\pgfmathprintnumber {8219.0860}}$\\%
Web-UK&\pgfutilensuremath {17{,}604.90}&\pgfutilensuremath {18{,}754.40}&$\mathbf {\pgfmathprintnumber[fixed,precision=2,zerofill] {12275.9}}$&--&\pgfutilensuremath {166{,}531.10}\\\bottomrule %
\end {tabular*}%

}
\vspace*{.25cm}
\subfloat[Index Size (Kb)]{
\begin {tabular*}{\textwidth}{@{\extracolsep{\fill}}rrrrrrr}%
\toprule \bfseries Dataset&\bfseries Ferrari-L&\bfseries Ferrari-G&\bfseries Grail&\bfseries Interval&\bfseries PWAH-8\\\midrule %
YAGO2&\pgfutilensuremath {372{,}150.70}&\pgfutilensuremath {448{,}139.06}&\pgfutilensuremath {575{,}701.28}&\pgfutilensuremath {182{,}962.96}&$\mathbf {\pgfmathprintnumber[fixed,precision=2,zerofill] {137878.214844}}$\\%
GovWild&$\mathbf {\pgfmathprintnumber[fixed,precision=2,zerofill] {206475.057617}}$&\pgfutilensuremath {297{,}724.03}&\pgfutilensuremath {282{,}054.38}&\pgfutilensuremath {921{,}605.13}&\pgfutilensuremath {311{,}359.35}\\%
Twitter&\pgfutilensuremath {384{,}049.21}&\pgfutilensuremath {384{,}368.44}&\pgfutilensuremath {637{,}072.31}&$\mathbf {\pgfmathprintnumber[fixed,precision=2,zerofill] {85648.1171875}}$&\pgfutilensuremath {97{,}859.81}\\%
Web-UK&\pgfutilensuremath {616{,}486.63}&\pgfutilensuremath {647{,}050.45}&\pgfutilensuremath {799{,}932.80}&--&$\mathbf {\pgfmathprintnumber[fixed,precision=2,zerofill] {266342.828125}}$\\\bottomrule %
\end {tabular*}%

}
\vspace*{.25cm}
\subfloat[Query Processing Performance (ms), 100k random queries]{
\begin {tabular*}{\textwidth}{@{\extracolsep{\fill}}rrrrrrr}%
\toprule \bfseries Dataset&\bfseries Ferrari-L&\bfseries Ferrari-G&\bfseries Grail&\bfseries Interval&\bfseries PWAH-8\\\midrule %
YAGO2&\pgfutilensuremath {12.00}&\pgfutilensuremath {10.95}&\pgfutilensuremath {16.56}&$\mathbf {\pgfmathprintnumber {10.4500}}$&\pgfutilensuremath {12.62}\\%
GovWild&\pgfutilensuremath {60.27}&\pgfutilensuremath {31.77}&\pgfutilensuremath {42.62}&$\mathbf {\pgfmathprintnumber {13.3260}}$&\pgfutilensuremath {33.30}\\%
Twitter&$\mathbf {\pgfmathprintnumber {5.552}}$&\pgfutilensuremath {5.65}&\pgfutilensuremath {19.27}&\pgfutilensuremath {8.66}&\pgfutilensuremath {10.32}\\%
Web-UK&$\mathbf {\pgfmathprintnumber {19.113}}$&\pgfutilensuremath {19.29}&\pgfutilensuremath {39.21}&--&\pgfutilensuremath {20.45}\\\bottomrule %
\end {tabular*}%

}
\vspace*{.25cm}
\subfloat[Query Processing Performance (ms), 100k positive queries]{
\begin {tabular*}{\textwidth}{@{\extracolsep{\fill}}rrrrrrr}%
\toprule \bfseries Dataset&\bfseries Ferrari-L&\bfseries Ferrari-G&\bfseries Grail&\bfseries Interval&\bfseries PWAH-8\\\midrule %
YAGO2&\pgfutilensuremath {59.39}&\pgfutilensuremath {38.43}&\pgfutilensuremath {97.99}&$\mathbf {\pgfmathprintnumber[fixed,precision=2,zerofill] {21.7020}}$&\pgfutilensuremath {44.19}\\%
GovWild&\pgfutilensuremath {171.46}&\pgfutilensuremath {85.12}&\pgfutilensuremath {228.98}&$\mathbf {\pgfmathprintnumber {29.8390}}$&\pgfutilensuremath {126.96}\\%
Twitter&\pgfutilensuremath {10.24}&$\mathbf {\pgfmathprintnumber {10.184}}$&\pgfutilensuremath {76.07}&\pgfutilensuremath {18.21}&\pgfutilensuremath {36.01}\\%
Web-UK&\pgfutilensuremath {25.54}&$\mathbf {\pgfmathprintnumber {18.007}}$&\pgfutilensuremath {95.25}&--&\pgfutilensuremath {43.73}\\\bottomrule %
\end {tabular*}%

}
\caption{Experimental Evaluation on Web Datasets}
\label{tab:exp-web}
\end{table}

As an additional note, the index size of Interval is sometimes smaller than \ferrari which seems to
be counterintuitive at first glance. The reason for this lies in the additional information
maintained at every node by \ferrari, for use in early pruning heuristics. In relatively sparse
datasets like Twitter and Yago2 the overhead of this extra information tends to outweigh the gains
made by interval merging. If needed, it is possible to turn off these heuristics easily to get
around this problem. However, we retain them across the board to avoid dataset-specific tuning of
the index.

\subsubsection{Query Processing} Finally, we turn our attention to the query processing performance
over Web-scale datasets. As the results summarized in Tables~\ref{tab:exp-web}c and
\ref{tab:exp-web}d demonstrate, the \ferrari variants and the Interval index provide the fastest
index structures. For Web-UK and Twitter, the \ferrari variants outperform all other approaches.  The
performance of GRAIL, as predicted, and PWAH-8 are inferior in comparison to Interval and \ferrari-G
when dealing with both random and positive query loads.

In summary, our experimental results indicate that \ferrari, in particular the global budgeted
variant, is highly scalable and consistent in being able to index a wide variety of graphs, and can
answer queries extremely fast\footnote{In a light-hearted view, our index shares the characteristics
  of its namesake F-1 racing car, in the sense that although not being the most powerful in the
  fray, it consistently tends to win on all circuits under diverse input conditions.}. This, we
believe, provides a compelling reason to use \ferrari-G on a wide spectrum of graph analytics
applications involving large to massive-scale graphs.


\section{Conclusions \& Outlook}
\label{sec:concl}
In this paper, we presented an efficient and scalable reachability index structure, \ferrari, that
allows to directly control the query processing/space consumption tradeoff via a user-specified
restriction on the resulting index size.  The two different variants of our index allow to either
specify the maximum size of the resulting node labels, or to impose a global size constraint which
allows the dynamic allocation of budgets based on the importance of individual nodes.  Using a
theoretically sound technique, \ferrari assigns a mixture of exact and approximate identifier ranges
to nodes so as to speed up both random as well as positive reachability queries. Using an extensive
array of experiments, we demonstrated that the resulting index can scale to massive-size graphs
quite easily, even when some of the state of the art indices fail to complete the
construction. \ferrari provides very fast query execution, demonstrating substantial gains in
processing time of both random and positive queries when compared to the previous state-of-the-art
method, GRAIL.

Results presented in this paper open up a range of possible future directions that we plan to
pursue. First of all, we would like to integrate \ferrari in a number of graph analytics algorithms,
starting from shortest path computation to more complex graph mining techniques (e.\,g.  Steiner
trees) in order to study its impact on the performance of these expensive operations. We also would
like to pursue further optimizations of the \ferrari index by, for instance, the use of an improved
tree covering algorithm designed to work synergistically with \ferrari to generate high quality
intervals to begin with, and the use of compression techniques for interval encoding afterwards.

\appendix
\section{Additional Experimental Results}
\subsection{Benchmark Datasets}
The complete results for our experimental evaluation over different size constraints are depicted in
Tables~\ref{tab:exp-benchmark-qp-vark} and \ref{tab:exp-benchmark-index-vark}.

\begin{sidewaystable}
\centering
\subfloat[Query Processing Performance (ms), 100k random queries]{
 \resizebox{\textwidth}{!}{
\begin {tabular}{rrrrrrrrrrrrrrrr}%
\toprule 
& \multicolumn{4}{c}{\bfseries Ferrari-L} & \multicolumn{4}{c}{\bfseries
  Ferrari-G} & \multicolumn{4}{c}{\bfseries Grail} 
& 
& 
& 
\\
\cmidrule(r){2-5} \cmidrule(r){6-9} \cmidrule(r){10-13}
\bfseries Dataset & 
\multicolumn{1}{c}{\footnotesize $k=1$} & \multicolumn{1}{c}{\footnotesize $k=2$} & \multicolumn{1}{c}{\footnotesize $k=3$} & \multicolumn{1}{c}{\footnotesize $k=5$} &
\multicolumn{1}{c}{\footnotesize $k=1$} & \multicolumn{1}{c}{\footnotesize $k=2$} & \multicolumn{1}{c}{\footnotesize $k=3$} & \multicolumn{1}{c}{\footnotesize $k=5$} &
\multicolumn{1}{c}{\footnotesize $d=1$} & \multicolumn{1}{c}{\footnotesize $d=2$} & \multicolumn{1}{c}{\footnotesize $d=3$} & \multicolumn{1}{c}{\footnotesize $d=5$} &
\bfseries PathTree & \bfseries Interval & \bfseries PWAH-8 \\
\midrule
ArXiV&\pgfutilensuremath {33.66}&\pgfutilensuremath {26.61}&\pgfutilensuremath {23.69}&\pgfutilensuremath {19.50}&\pgfutilensuremath {30.90}&\pgfutilensuremath {20.73}&\pgfutilensuremath {13.91}&\pgfutilensuremath {8.78}&\pgfutilensuremath {180.26}&\pgfutilensuremath {122.85}&\pgfutilensuremath {100.92}&\pgfutilensuremath {75.41}&$\mathbf {\pgfmathprintnumber {3.414}}$&\pgfutilensuremath {4.17}&\pgfutilensuremath {23.22}\\%
Pubmed&\pgfutilensuremath {10.20}&\pgfutilensuremath {8.23}&\pgfutilensuremath {7.58}&\pgfutilensuremath {6.05}&\pgfutilensuremath {7.66}&\pgfutilensuremath {5.77}&\pgfutilensuremath {4.88}&\pgfutilensuremath {3.73}&\pgfutilensuremath {20.64}&\pgfutilensuremath {13.87}&\pgfutilensuremath {12.27}&\pgfutilensuremath {10.55}&$\mathbf {\pgfmathprintnumber {2.758}}$&\pgfutilensuremath {3.16}&\pgfutilensuremath {28.58}\\%
Human&\pgfutilensuremath {0.78}&\pgfutilensuremath {0.78}&$\mathbf {\pgfmathprintnumber {0.76}}$&\pgfutilensuremath {0.77}&\pgfutilensuremath {0.77}&\pgfutilensuremath {0.78}&\pgfutilensuremath {0.78}&\pgfutilensuremath {0.77}&\pgfutilensuremath {4.80}&\pgfutilensuremath {4.98}&\pgfutilensuremath {5.04}&\pgfutilensuremath {5.29}&\pgfutilensuremath {1.21}&\pgfutilensuremath {1.07}&\pgfutilensuremath {1.06}\\%
GO&\pgfutilensuremath {6.89}&\pgfutilensuremath {4.45}&\pgfutilensuremath {4.10}&\pgfutilensuremath {3.51}&\pgfutilensuremath {4.28}&\pgfutilensuremath {3.41}&\pgfutilensuremath {2.96}&\pgfutilensuremath {2.85}&\pgfutilensuremath {10.33}&\pgfutilensuremath {6.46}&\pgfutilensuremath {4.83}&\pgfutilensuremath {4.13}&$\mathbf {\pgfmathprintnumber {2.038}}$&\pgfutilensuremath {2.47}&\pgfutilensuremath {4.45}\\%
CiteSeer&\pgfutilensuremath {8.02}&\pgfutilensuremath {6.57}&\pgfutilensuremath {6.40}&\pgfutilensuremath {6.13}&\pgfutilensuremath {6.14}&\pgfutilensuremath {6.18}&\pgfutilensuremath {6.15}&\pgfutilensuremath {6.24}&\pgfutilensuremath {8.54}&\pgfutilensuremath {8.28}&\pgfutilensuremath {8.19}&\pgfutilensuremath {8.05}&$\mathbf {\pgfmathprintnumber {5.011}}$&\pgfutilensuremath {8.28}&\pgfutilensuremath {12.39}\\%
CiteSeerX&\pgfutilensuremath {20.71}&\pgfutilensuremath {17.90}&\pgfutilensuremath {17.04}&\pgfutilensuremath {15.88}&\pgfutilensuremath {16.09}&\pgfutilensuremath {12.62}&\pgfutilensuremath {10.95}&\pgfutilensuremath {9.31}&\pgfutilensuremath {63.06}&\pgfutilensuremath {50.02}&\pgfutilensuremath {46.49}&\pgfutilensuremath {41.23}&--&$\mathbf {\pgfmathprintnumber {9.274}}$&\pgfutilensuremath {21.32}\\%
GO-Uniprot&\pgfutilensuremath {32.45}&\pgfutilensuremath {28.92}&\pgfutilensuremath
{29.69}&\pgfutilensuremath {28.30}&\pgfutilensuremath {32.27}&\pgfutilensuremath
{29.36}&\pgfutilensuremath {28.79}&\pgfutilensuremath {28.92}&\pgfutilensuremath
{6.51}&\pgfutilensuremath {6.01}&\pgfutilensuremath {6.01}&$\mathbf {\pgfmathprintnumber {5.94}}$&--&\pgfutilensuremath {16.82}&\pgfutilensuremath {48.70}\\%
Cit-Patents&\pgfutilensuremath {1{,}100.32}&\pgfutilensuremath {858.15}&\pgfutilensuremath
{798.86}&\pgfutilensuremath {778.09}&\pgfutilensuremath {900.56}&\pgfutilensuremath
{722.61}&\pgfutilensuremath {672.76}&$\mathbf{ \pgfmathprintnumber[precision=2,fixed,zerofill] {502.20}}$ &\pgfutilensuremath {2{,}545.37}&\pgfutilensuremath {1{,}123.68}&\pgfutilensuremath {851.24}&\pgfutilensuremath {578.83}&--&--&\pgfutilensuremath {1{,}514.91}\\\bottomrule %
\end {tabular}%
}

}
\vspace*{.5cm}
 \subfloat[Query Processing Performance (ms), 100k positive queries]{
\resizebox{\textwidth}{!}{
\begin {tabular}{rrrrrrrrrrrrrrrr}%
\toprule 
& \multicolumn{4}{c}{\bfseries Ferrari-L} & \multicolumn{4}{c}{\bfseries
  Ferrari-G} & \multicolumn{4}{c}{\bfseries Grail} 
& 
& 
& 
\\
\cmidrule(r){2-5} \cmidrule(r){6-9} \cmidrule(r){10-13}
\bfseries Dataset & 
\multicolumn{1}{c}{\footnotesize $k=1$} & \multicolumn{1}{c}{\footnotesize $k=2$} & \multicolumn{1}{c}{\footnotesize $k=3$} & \multicolumn{1}{c}{\footnotesize $k=5$} &
\multicolumn{1}{c}{\footnotesize $k=1$} & \multicolumn{1}{c}{\footnotesize $k=2$} & \multicolumn{1}{c}{\footnotesize $k=3$} & \multicolumn{1}{c}{\footnotesize $k=5$} &
\multicolumn{1}{c}{\footnotesize $d=1$} & \multicolumn{1}{c}{\footnotesize $d=2$} & \multicolumn{1}{c}{\footnotesize $d=3$} & \multicolumn{1}{c}{\footnotesize $d=5$} &
\bfseries PathTree & \bfseries Interval & \bfseries PWAH-8 \\
\midrule
ArXiV&\pgfutilensuremath {86.89}&\pgfutilensuremath {72.11}&\pgfutilensuremath {62.64}&\pgfutilensuremath {52.77}&\pgfutilensuremath {84.12}&\pgfutilensuremath {56.85}&\pgfutilensuremath {37.98}&\pgfutilensuremath {21.30}&\pgfutilensuremath {398.56}&\pgfutilensuremath {275.74}&\pgfutilensuremath {220.31}&\pgfutilensuremath {172.43}&$\mathbf {\pgfmathprintnumber {4.937}}$&\pgfutilensuremath {5.95}&\pgfutilensuremath {17.74}\\%
Pubmed&\pgfutilensuremath {39.17}&\pgfutilensuremath {33.38}&\pgfutilensuremath {31.31}&\pgfutilensuremath {25.83}&\pgfutilensuremath {31.20}&\pgfutilensuremath {24.26}&\pgfutilensuremath {20.28}&\pgfutilensuremath {12.09}&\pgfutilensuremath {123.42}&\pgfutilensuremath {91.34}&\pgfutilensuremath {85.38}&\pgfutilensuremath {79.78}&$\mathbf {\pgfmathprintnumber {4.415}}$&\pgfutilensuremath {6.21}&\pgfutilensuremath {43.58}\\%
Human&\pgfutilensuremath {2.48}&\pgfutilensuremath {2.08}&\pgfutilensuremath {2.04}&\pgfutilensuremath {2.02}&\pgfutilensuremath {2.03}&\pgfutilensuremath {1.96}&\pgfutilensuremath {2.02}&\pgfutilensuremath {2.04}&\pgfutilensuremath {14.16}&\pgfutilensuremath {14.48}&\pgfutilensuremath {15.22}&\pgfutilensuremath {18.25}&\pgfutilensuremath {1.30}&\pgfutilensuremath {1.79}&\pgfutilensuremath {6.07}\\%
GO&\pgfutilensuremath {12.89}&\pgfutilensuremath {10.70}&\pgfutilensuremath {10.72}&\pgfutilensuremath {6.92}&\pgfutilensuremath {10.05}&\pgfutilensuremath {5.80}&\pgfutilensuremath {4.64}&\pgfutilensuremath {4.25}&\pgfutilensuremath {24.77}&\pgfutilensuremath {20.06}&\pgfutilensuremath {19.59}&\pgfutilensuremath {19.84}&$\mathbf {\pgfmathprintnumber {2.044}}$&\pgfutilensuremath {3.26}&\pgfutilensuremath {11.43}\\%
CiteSeer&\pgfutilensuremath {25.23}&\pgfutilensuremath {18.96}&\pgfutilensuremath {14.33}&\pgfutilensuremath {13.37}&\pgfutilensuremath {13.52}&\pgfutilensuremath {13.33}&\pgfutilensuremath {13.51}&\pgfutilensuremath {13.47}&\pgfutilensuremath {72.81}&\pgfutilensuremath {77.57}&\pgfutilensuremath {83.83}&\pgfutilensuremath {85.22}&$\mathbf {\pgfmathprintnumber {6.119}}$&\pgfutilensuremath {15.17}&\pgfutilensuremath {30.60}\\%
CiteSeerX&\pgfutilensuremath {106.56}&\pgfutilensuremath {96.52}&\pgfutilensuremath {102.09}&\pgfutilensuremath {82.76}&\pgfutilensuremath {86.80}&\pgfutilensuremath {64.20}&\pgfutilensuremath {53.04}&\pgfutilensuremath {43.06}&\pgfutilensuremath {882.49}&\pgfutilensuremath {731.03}&\pgfutilensuremath {741.47}&\pgfutilensuremath {700.49}&--&$\mathbf {\pgfmathprintnumber {30.382}}$&\pgfutilensuremath {69.21}\\%
GO-Uniprot&\pgfutilensuremath {68.90}&\pgfutilensuremath {71.62}&\pgfutilensuremath {66.51}&\pgfutilensuremath {65.00}&\pgfutilensuremath {71.70}&\pgfutilensuremath {73.24}&\pgfutilensuremath {67.69}&\pgfutilensuremath {64.72}&\pgfutilensuremath {135.06}&\pgfutilensuremath {125.52}&\pgfutilensuremath {126.90}&\pgfutilensuremath {131.46}&--&$\mathbf{\pgfmathprintnumber {31.76}}$&\pgfutilensuremath {54.55}\\%
Cit-Patents&\pgfutilensuremath {6{,}581.21}&\pgfutilensuremath {4{,}836.69}&\pgfutilensuremath {4{,}443.29}&\pgfutilensuremath {4{,}086.21}&\pgfutilensuremath {5{,}210.22}&\pgfutilensuremath {4{,}016.88}&\pgfutilensuremath {3{,}621.21}&\pgfutilensuremath {2{,}667.38}&\pgfutilensuremath {15{,}743.66}&\pgfutilensuremath {9{,}865.95}&\pgfutilensuremath {7{,}712.84}&\pgfutilensuremath {5{,}409.82}&--&--&$\mathbf{\pgfmathprintnumber[precision=2,fixed,zerofill] {1739.30}}$\\\bottomrule %
\end {tabular}%
}

}
\caption{Experimental Evaluation on Benchmark Datasets -- Query Processing (Varying Size Constraint)}
\label{tab:exp-benchmark-qp-vark}
\end{sidewaystable}

\begin{sidewaystable}
\centering
\subfloat[Index Size (Kb)]{
\resizebox{\textwidth}{!}{
\begin {tabular}{rrrrrrrrrrrrrrrr}%
\toprule 
& \multicolumn{4}{c}{\bfseries Ferrari-L} & \multicolumn{4}{c}{\bfseries
  Ferrari-G} & \multicolumn{4}{c}{\bfseries Grail} 
& 
& 
& 
\\
\cmidrule(r){2-5} \cmidrule(r){6-9} \cmidrule(r){10-13}
\bfseries Dataset & 
\multicolumn{1}{c}{\footnotesize $k=1$} & \multicolumn{1}{c}{\footnotesize $k=2$} & \multicolumn{1}{c}{\footnotesize $k=3$} & \multicolumn{1}{c}{\footnotesize $k=5$} &
\multicolumn{1}{c}{\footnotesize $k=1$} & \multicolumn{1}{c}{\footnotesize $k=2$} & \multicolumn{1}{c}{\footnotesize $k=3$} & \multicolumn{1}{c}{\footnotesize $k=5$} &
\multicolumn{1}{c}{\footnotesize $d=1$} & \multicolumn{1}{c}{\footnotesize $d=2$} & \multicolumn{1}{c}{\footnotesize $d=3$} & \multicolumn{1}{c}{\footnotesize $d=5$} &
\bfseries PathTree & \bfseries Interval & \bfseries PWAH-8 \\
\midrule
ArXiV&\pgfutilensuremath {164.44}&\pgfutilensuremath {205.17}&\pgfutilensuremath {243.86}&\pgfutilensuremath {312.14}&\pgfutilensuremath {169.91}&\pgfutilensuremath {222.63}&\pgfutilensuremath {275.33}&\pgfutilensuremath {380.86}&$\mathbf{\pgfmathprintnumber {117.19}}$&\pgfutilensuremath {210.94}&\pgfutilensuremath {304.69}&\pgfutilensuremath {492.19}&$\pgfmathprintnumber[precision=2,fixed,zerofill] {338.07}$&\pgfutilensuremath {1{,}364.99}&\pgfutilensuremath {315.24}\\%
Pubmed&\pgfutilensuremath {213.56}&\pgfutilensuremath {250.32}&\pgfutilensuremath
{283.68}&\pgfutilensuremath {340.43}&\pgfutilensuremath {254.88}&\pgfutilensuremath
{333.98}&\pgfutilensuremath {413.06}&\pgfutilensuremath {571.15}&$\mathbf{ \pgfmathprintnumber {175.78}}$&\pgfutilensuremath {316.41}&\pgfutilensuremath {457.03}&\pgfutilensuremath {738.28}&${\pgfmathprintnumber[precision=2,fixed] {419.03}}$&\pgfutilensuremath {1{,}523.83}&\pgfutilensuremath {358.96}\\%
Human&\pgfutilensuremath {764.38}&\pgfutilensuremath {768.88}&\pgfutilensuremath
{769.61}&\pgfutilensuremath {769.95}&\pgfutilensuremath {769.85}&\pgfutilensuremath
{770.30}&\pgfutilensuremath {770.75}&\pgfutilensuremath {770.78}&\pgfutilensuremath
{758.03}&\pgfutilensuremath {1{,}364.45}&\pgfutilensuremath {1{,}970.87}&\pgfutilensuremath
{3{,}183.71}&${ \pgfmathprintnumber[fixed,precision=2,zerofill] {458.01}}$&\pgfutilensuremath
{160.56}&$\mathbf {\pgfmathprintnumber {160.22}}$\\%
GO&\pgfutilensuremath {165.25}&\pgfutilensuremath {186.05}&\pgfutilensuremath
{200.37}&\pgfutilensuremath {217.45}&\pgfutilensuremath {192.36}&\pgfutilensuremath
{231.57}&\pgfutilensuremath {251.01}&\pgfutilensuremath {265.21}&\pgfutilensuremath
{132.68}&\pgfutilensuremath {238.82}&\pgfutilensuremath {344.96}&\pgfutilensuremath {557.24}&$
{\pgfmathprintnumber[fixed,precision=2,zerofill] {133.30}}$&\pgfutilensuremath {180.58}&$\mathbf {\pgfmathprintnumber {81.86}}$\\%
CiteSeer&\pgfutilensuremath {10{,}876.86}&\pgfutilensuremath {13{,}220.50}&\pgfutilensuremath
{13{,}844.14}&\pgfutilensuremath {13{,}933.90}&\pgfutilensuremath {13{,}927.15}&\pgfutilensuremath
{13{,}934.29}&\pgfutilensuremath {13{,}934.29}&\pgfutilensuremath {13{,}934.29}&\pgfutilensuremath
{13{,}553.65}&\pgfutilensuremath {24{,}396.57}&\pgfutilensuremath {35{,}239.50}&\pgfutilensuremath
{56{,}925.34}&$ {\pgfmathprintnumber {9221.61}}$&\pgfutilensuremath {7{,}733.94}& $\mathbf {\pgfmathprintnumber {6723.36}}$\\%
CiteSeerX&\pgfutilensuremath {134{,}770.72}&\pgfutilensuremath {140{,}793.14}&\pgfutilensuremath
{146{,}975.22}&\pgfutilensuremath {158{,}046.72}&\pgfutilensuremath
{152{,}661.80}&\pgfutilensuremath {187{,}058.08}&\pgfutilensuremath
{209{,}991.64}&\pgfutilensuremath {242{,}236.08}&$\mathbf{\pgfmathprintnumber[fixed,precision=2,zerofill]
{127742.21}}$&\pgfutilensuremath {229{,}935.97}&\pgfutilensuremath
{332{,}129.74}&\pgfutilensuremath
{536{,}517.27}&--&${\pgfmathprintnumber[fixed,precision=2,zerofill] {430913.355469}}$& ${\pgfmathprintnumber[fixed] {152354.44}}$\\%
GO-Uniprot&\pgfutilensuremath {197{,}329.62}&\pgfutilensuremath {258{,}566.04}&\pgfutilensuremath {318{,}769.95}&\pgfutilensuremath {429{,}564.04}&\pgfutilensuremath {197{,}334.69}&\pgfutilensuremath {258{,}576.47}&\pgfutilensuremath {319{,}818.29}&\pgfutilensuremath {442{,}301.79}&$\mathbf{\pgfmathprintnumber[fixed,precision=2,zerofill] {136092.89}}$&\pgfutilensuremath {244{,}967.20}&\pgfutilensuremath {353{,}841.52}&\pgfutilensuremath {571{,}590.14}&--&\pgfutilensuremath {774{,}081.33}&\pgfutilensuremath {249{,}883.80}\\%
Cit-Patents&\pgfutilensuremath {92{,}089.32}&\pgfutilensuremath {105{,}895.74}&\pgfutilensuremath
{121{,}532.11}&\pgfutilensuremath {151{,}631.73}&\pgfutilensuremath
{106{,}902.60}&\pgfutilensuremath {140{,}079.25}&\pgfutilensuremath
{173{,}255.88}&\pgfutilensuremath {239{,}609.23}&$\mathbf{ \pgfmathprintnumber {73725.94}}$&\pgfutilensuremath {132{,}706.69}&\pgfutilensuremath {191{,}687.44}&\pgfutilensuremath {309{,}648.94}&--&--&\pgfutilensuremath {5{,}462{,}135.76}\\\bottomrule %
\end {tabular}%
}

}
\vspace*{.5cm}
 \subfloat[Construction Time (ms)]{
\resizebox{\textwidth}{!}{
\begin {tabular}{rrrrrrrrrrrrrrrr}%
\toprule 
& \multicolumn{4}{c}{\bfseries Ferrari-L} & \multicolumn{4}{c}{\bfseries
  Ferrari-G} & \multicolumn{4}{c}{\bfseries Grail} 
& 
& 
\\
\cmidrule(r){2-5} \cmidrule(r){6-9} \cmidrule(r){10-13}
\bfseries Dataset & 
\multicolumn{1}{c}{\footnotesize $k=1$} & \multicolumn{1}{c}{\footnotesize $k=2$} & \multicolumn{1}{c}{\footnotesize $k=3$} & \multicolumn{1}{c}{\footnotesize $k=5$} &
\multicolumn{1}{c}{\footnotesize $k=1$} & \multicolumn{1}{c}{\footnotesize $k=2$} & \multicolumn{1}{c}{\footnotesize $k=3$} & \multicolumn{1}{c}{\footnotesize $k=5$} &
\multicolumn{1}{c}{\footnotesize $d=1$} & \multicolumn{1}{c}{\footnotesize $d=2$} & \multicolumn{1}{c}{\footnotesize $d=3$} & \multicolumn{1}{c}{\footnotesize $d=5$} &
\bfseries PathTree & \bfseries Interval & \bfseries PWAH-8 \\
\midrule
ArXiV&\pgfutilensuremath {7.72}&\pgfutilensuremath {11.86}&\pgfutilensuremath {15.84}&\pgfutilensuremath {19.49}&\pgfutilensuremath {14.86}&\pgfutilensuremath {22.15}&\pgfutilensuremath {26.62}&\pgfutilensuremath {33.50}&$\mathbf{\pgfmathprintnumber {2.59}}$&\pgfutilensuremath {4.57}&\pgfutilensuremath {7.86}&\pgfutilensuremath {14.03}&$\pgfmathprintnumber {4537.39}$&\pgfutilensuremath {34.54}&\pgfutilensuremath {70.10}\\%
Pubmed&\pgfutilensuremath {10.56}&\pgfutilensuremath {13.09}&\pgfutilensuremath
{14.28}&\pgfutilensuremath {17.67}&\pgfutilensuremath {16.42}&\pgfutilensuremath
{23.31}&\pgfutilensuremath {24.54}&\pgfutilensuremath {27.11}&$\mathbf{ \pgfmathprintnumber[fixed,precision=2,zerofill] {3.00}}$&\pgfutilensuremath {5.21}&\pgfutilensuremath {8.21}&\pgfutilensuremath {14.13}&$\pgfmathprintnumber {326.541}$&\pgfutilensuremath {20.35}&\pgfutilensuremath {44.41}\\%
Human&\pgfutilensuremath {23.71}&\pgfutilensuremath {23.36}&\pgfutilensuremath
{22.77}&\pgfutilensuremath {23.97}&\pgfutilensuremath {23.89}&\pgfutilensuremath
{23.37}&\pgfutilensuremath {22.90}&\pgfutilensuremath {23.33}&\pgfutilensuremath
{10.08}&\pgfutilensuremath {15.93}&\pgfutilensuremath {32.72}&\pgfutilensuremath
{64.36}&\pgfutilensuremath {348.48}&$\mathbf{\pgfmathprintnumber[fixed,precision=2,zerofill] {2.70}}$&\pgfutilensuremath {3.82}\\%
GO&\pgfutilensuremath {5.34}&\pgfutilensuremath {6.03}&\pgfutilensuremath {6.48}&\pgfutilensuremath
{6.62}&\pgfutilensuremath {6.74}&\pgfutilensuremath {6.72}&\pgfutilensuremath
{6.91}&\pgfutilensuremath {6.94}&$\mathbf{ \pgfmathprintnumber {1.93}}$&\pgfutilensuremath {3.23}&\pgfutilensuremath {4.83}&\pgfutilensuremath {7.77}&$\pgfmathprintnumber {89.834}$&\pgfutilensuremath {5.06}&\pgfutilensuremath {8.67}\\%
CiteSeer&\pgfutilensuremath {483.24}&\pgfutilensuremath {478.34}&\pgfutilensuremath
{455.58}&\pgfutilensuremath {450.12}&\pgfutilensuremath {477.39}&\pgfutilensuremath
{463.38}&\pgfutilensuremath {459.42}&\pgfutilensuremath {459.90}&\pgfutilensuremath
{273.73}&\pgfutilensuremath {487.09}&\pgfutilensuremath {1{,}037.21}&\pgfutilensuremath
{2{,}015.90}&$\pgfmathprintnumber {26479.7}$& $\mathbf{ \pgfmathprintnumber[precision=2,zerofill] {251.10}}$&\pgfutilensuremath {416.41}\\%
CiteSeerX&\pgfutilensuremath {12{,}869.70}&\pgfutilensuremath {13{,}336.30}&\pgfutilensuremath {13{,}748.50}&\pgfutilensuremath {14{,}110.20}&\pgfutilensuremath {13{,}978.20}&\pgfutilensuremath {14{,}970.80}&\pgfutilensuremath {15{,}415.50}&\pgfutilensuremath {16{,}233.40}&\pgfutilensuremath {3{,}397.65}&\pgfutilensuremath {6{,}638.96}&\pgfutilensuremath {11{,}436.70}&\pgfutilensuremath {20{,}528.40}&--&$\mathbf {\pgfmathprintnumber {5808.787}}$&\pgfutilensuremath {14{,}444.09}\\%
GO-Uniprot&\pgfutilensuremath {9{,}173.19}&\pgfutilensuremath {15{,}526.80}&\pgfutilensuremath
{19{,}205.90}&\pgfutilensuremath {26{,}105.90}&\pgfutilensuremath {15{,}495.80}&\pgfutilensuremath
{23{,}040.90}&\pgfutilensuremath {26{,}817.10}&\pgfutilensuremath {29{,}611.90}&$\mathbf{ \pgfmathprintnumber {2841.04}}$&\pgfutilensuremath {5{,}131.31}&\pgfutilensuremath {14{,}764.30}&\pgfutilensuremath {34{,}518.40}&--&\pgfutilensuremath {15{,}213.55}&\pgfutilensuremath {26{,}745.61}\\%
Cit-Patents&\pgfutilensuremath {16{,}085.90}&\pgfutilensuremath {17{,}418.30}&\pgfutilensuremath
{18{,}649.30}&\pgfutilensuremath {20{,}665.50}&\pgfutilensuremath {20{,}211.80}&\pgfutilensuremath
{23{,}861.60}&\pgfutilensuremath {27{,}179.20}&\pgfutilensuremath {32{,}366.20}&$\mathbf{ \pgfmathprintnumber {3842.96}}$&\pgfutilensuremath {7{,}516.79}&\pgfutilensuremath {12{,}123.50}&\pgfutilensuremath {21{,}621.70}&--&--&\pgfutilensuremath {751{,}984.08}\\\bottomrule %
\end {tabular}%
}
}
\caption{Experimental Evaluation on Benchmark Datasets -- Indexing (Varying Size Constraint)}
\label{tab:exp-benchmark-index-vark}
\end{sidewaystable}

\subsection{Web Datasets}
The complete results for our experimental evaluation over different size constraints are depicted in
Tables~\ref{tab:exp-web-qp-vark} and \ref{tab:exp-web-index-vark}.

\begin{sidewaystable}
\centering
\subfloat[Query Processing Performance (ms), 100k random queries]{
 \resizebox{\textwidth}{!}{
\begin {tabular}{rrrrrrrrrrrrrrr}%
\toprule 
& \multicolumn{4}{c}{\bfseries Ferrari-L} & \multicolumn{4}{c}{\bfseries
  Ferrari-G} & \multicolumn{4}{c}{\bfseries Grail} 
& 
& 
\\
\cmidrule(r){2-5} \cmidrule(r){6-9} \cmidrule(r){10-13}
\bfseries Dataset & 
\multicolumn{1}{c}{\footnotesize $k=1$} & \multicolumn{1}{c}{\footnotesize $k=2$} & \multicolumn{1}{c}{\footnotesize $k=5$} & \multicolumn{1}{c}{\footnotesize $k=10$} &
\multicolumn{1}{c}{\footnotesize $k=1$} & \multicolumn{1}{c}{\footnotesize $k=2$} & \multicolumn{1}{c}{\footnotesize $k=5$} & \multicolumn{1}{c}{\footnotesize $k=10$} &
\multicolumn{1}{c}{\footnotesize $d=1$} & \multicolumn{1}{c}{\footnotesize $d=2$} & \multicolumn{1}{c}{\footnotesize $d=5$} & \multicolumn{1}{c}{\footnotesize $d=10$} &
\bfseries Interval & \bfseries PWAH-8 \\
\midrule
YAGO2&\pgfutilensuremath {14.82}&\pgfutilensuremath {12.00}&\pgfutilensuremath
{11.19}&\pgfutilensuremath {10.98}&\pgfutilensuremath {11.13}&\pgfutilensuremath
{10.95}&\pgfutilensuremath {10.73}&\pgfutilensuremath {10.78}&\pgfutilensuremath
{30.53}&\pgfutilensuremath {16.56}&\pgfutilensuremath {16.66}&\pgfutilensuremath {16.80}&$\mathbf {\pgfmathprintnumber {10.45}}$&\pgfutilensuremath {12.62}\\%
GovWild&\pgfutilensuremath {102.29}&\pgfutilensuremath {60.27}&\pgfutilensuremath
{51.70}&\pgfutilensuremath {36.62}&\pgfutilensuremath {78.64}&\pgfutilensuremath
{31.77}&\pgfutilensuremath {21.91}&\pgfutilensuremath {20.38}&\pgfutilensuremath
{76.04}&\pgfutilensuremath {42.62}&\pgfutilensuremath {33.53}&\pgfutilensuremath {32.98}&$\mathbf{ \pgfmathprintnumber {13.33}}$&\pgfutilensuremath {33.30}\\%
Twitter&\pgfutilensuremath {6.12}&$\mathbf{ \pgfmathprintnumber {5.55}}$&\pgfutilensuremath {5.58}&\pgfutilensuremath {5.68}&\pgfutilensuremath {5.60}&\pgfutilensuremath {5.65}&\pgfutilensuremath {5.69}&\pgfutilensuremath {5.71}&\pgfutilensuremath {18.76}&\pgfutilensuremath {19.27}&\pgfutilensuremath {19.96}&\pgfutilensuremath {21.11}&\pgfutilensuremath {8.66}&\pgfutilensuremath {10.32}\\%
Web-UK&\pgfutilensuremath {22.74}&\pgfutilensuremath {19.11}&\pgfutilensuremath
{19.25}&\pgfutilensuremath {19.44}&\pgfutilensuremath {19.15}&\pgfutilensuremath {19.29}&$\mathbf{
  \pgfmathprintnumber {18.88}}$ &\pgfutilensuremath {18.92}&\pgfutilensuremath {42.24}&\pgfutilensuremath {39.21}&\pgfutilensuremath {40.53}&\pgfutilensuremath {43.00}&--&\pgfutilensuremath {20.45}\\\bottomrule %
\end {tabular}%
}
}
\vspace*{.5cm}
 \subfloat[Query Processing Performance (ms), 100k positive queries]{
 \resizebox{\textwidth}{!}{
\begin {tabular}{rrrrrrrrrrrrrrr}%
\toprule 
& \multicolumn{4}{c}{\bfseries Ferrari-L} & \multicolumn{4}{c}{\bfseries
  Ferrari-G} & \multicolumn{4}{c}{\bfseries Grail} 
& 
& 
\\
\cmidrule(r){2-5} \cmidrule(r){6-9} \cmidrule(r){10-13}
\bfseries Dataset & 
\multicolumn{1}{c}{\footnotesize $k=1$} & \multicolumn{1}{c}{\footnotesize $k=2$} & \multicolumn{1}{c}{\footnotesize $k=5$} & \multicolumn{1}{c}{\footnotesize $k=10$} &
\multicolumn{1}{c}{\footnotesize $k=1$} & \multicolumn{1}{c}{\footnotesize $k=2$} & \multicolumn{1}{c}{\footnotesize $k=5$} & \multicolumn{1}{c}{\footnotesize $k=10$} &
\multicolumn{1}{c}{\footnotesize $d=1$} & \multicolumn{1}{c}{\footnotesize $d=2$} & \multicolumn{1}{c}{\footnotesize $d=5$} & \multicolumn{1}{c}{\footnotesize $d=10$} &
\bfseries Interval & \bfseries PWAH-8 \\
\midrule
YAGO2&\pgfutilensuremath {73.65}&\pgfutilensuremath {59.39}&\pgfutilensuremath {44.67}&\pgfutilensuremath {38.83}&\pgfutilensuremath {48.35}&\pgfutilensuremath {38.43}&\pgfutilensuremath {38.82}&\pgfutilensuremath {38.07}&\pgfutilensuremath {107.33}&\pgfutilensuremath {97.99}&\pgfutilensuremath {101.97}&\pgfutilensuremath {107.55}&$\mathbf{
  \pgfmathprintnumber[precision=2,zerofill,fixed] {21.70}}$&\pgfutilensuremath {44.19}\\%
GovWild&\pgfutilensuremath {293.99}&\pgfutilensuremath {171.46}&\pgfutilensuremath {124.11}&\pgfutilensuremath {75.04}&\pgfutilensuremath {166.02}&\pgfutilensuremath {85.12}&\pgfutilensuremath {45.72}&\pgfutilensuremath {42.51}&\pgfutilensuremath {248.24}&\pgfutilensuremath {228.98}&\pgfutilensuremath {165.33}&\pgfutilensuremath {167.25}&$\mathbf{
  \pgfmathprintnumber[precision=2,zerofill,fixed] {29.84}}$&\pgfutilensuremath {126.96}\\%
Twitter&\pgfutilensuremath {11.10}&\pgfutilensuremath {10.24}&$\mathbf{
  \pgfmathprintnumber[precision=2,zerofill,fixed] {9.80}}$&\pgfutilensuremath {11.63}&\pgfutilensuremath {10.34}&\pgfutilensuremath {10.18}&\pgfutilensuremath {10.18}&\pgfutilensuremath {10.08}&\pgfutilensuremath {248{,}494.31}&\pgfutilensuremath {76.07}&\pgfutilensuremath {83.57}&\pgfutilensuremath {96.13}&\pgfutilensuremath {18.21}&\pgfutilensuremath {36.01}\\%
Web-UK&\pgfutilensuremath {28.45}&\pgfutilensuremath {25.54}&\pgfutilensuremath {22.82}&\pgfutilensuremath {18.22}&\pgfutilensuremath {23.64}&\pgfutilensuremath {18.01}&\pgfutilensuremath {17.49}&$\mathbf{
  \pgfmathprintnumber[precision=2,zerofill,fixed] {16.85}}$ &\pgfutilensuremath {26{,}792.72}&\pgfutilensuremath {95.25}&\pgfutilensuremath {102.59}&\pgfutilensuremath {113.83}&--&\pgfutilensuremath {43.73}\\\bottomrule %
\end {tabular}%
}

}
\caption{Experimental Evaluation on Web Datasets -- Query Processing (Varying Size Constraint)}
\label{tab:exp-web-qp-vark}
\end{sidewaystable}

\begin{sidewaystable}
\centering
\subfloat[Index Size (Kb)]{
 \resizebox{\textwidth}{!}{
\begin {tabular}{rrrrrrrrrrrrrrr}%
\toprule 
& \multicolumn{4}{c}{\bfseries Ferrari-L} & \multicolumn{4}{c}{\bfseries
  Ferrari-G} & \multicolumn{4}{c}{\bfseries Grail} 
& 
& 
\\
\cmidrule(r){2-5} \cmidrule(r){6-9} \cmidrule(r){10-13}
\bfseries Dataset & 
\multicolumn{1}{c}{\footnotesize $k=1$} & \multicolumn{1}{c}{\footnotesize $k=2$} & \multicolumn{1}{c}{\footnotesize $k=5$} & \multicolumn{1}{c}{\footnotesize $k=10$} &
\multicolumn{1}{c}{\footnotesize $k=1$} & \multicolumn{1}{c}{\footnotesize $k=2$} & \multicolumn{1}{c}{\footnotesize $k=5$} & \multicolumn{1}{c}{\footnotesize $k=10$} &
\multicolumn{1}{c}{\footnotesize $d=1$} & \multicolumn{1}{c}{\footnotesize $d=2$} & \multicolumn{1}{c}{\footnotesize $d=5$} & \multicolumn{1}{c}{\footnotesize $d=10$} &
\bfseries Interval & \bfseries PWAH-8 \\
\midrule
YAGO2&\pgfutilensuremath {346{,}225.04}&\pgfutilensuremath {372{,}150.70}&\pgfutilensuremath {417{,}050.40}&\pgfutilensuremath {444{,}546.26}&\pgfutilensuremath {405{,}498.36}&\pgfutilensuremath {448{,}139.06}&\pgfutilensuremath {453{,}340.72}&\pgfutilensuremath {453{,}560.94}&\pgfutilensuremath {319{,}834.04}&\pgfutilensuremath {575{,}701.28}&\pgfutilensuremath {1{,}343{,}302.98}&--&\pgfutilensuremath {182{,}962.96}&$\mathbf{
  \pgfmathprintnumber[precision=2,zerofill,fixed] {137878.21}}$\\%
GovWild&\pgfutilensuremath {181{,}599.94}&\pgfutilensuremath {206{,}475.06}&\pgfutilensuremath {269{,}511.88}&\pgfutilensuremath {343{,}819.93}&\pgfutilensuremath {227{,}210.46}&\pgfutilensuremath {297{,}724.03}&\pgfutilensuremath {475{,}154.42}&\pgfutilensuremath {511{,}974.72}&$\mathbf{
  \pgfmathprintnumber[precision=2,zerofill,fixed] {156696.88}}$&\pgfutilensuremath {282{,}054.38}&\pgfutilensuremath {658{,}126.88}&\pgfutilensuremath {1{,}284{,}914.38}&\pgfutilensuremath {921{,}605.13}&\pgfutilensuremath {311{,}359.35}\\%
Twitter&\pgfutilensuremath {369{,}201.71}&\pgfutilensuremath {384{,}049.21}&\pgfutilensuremath {384{,}358.04}&\pgfutilensuremath {384{,}367.92}&\pgfutilensuremath {384{,}348.90}&\pgfutilensuremath {384{,}368.44}&\pgfutilensuremath {384{,}368.70}&\pgfutilensuremath {384{,}368.70}&\pgfutilensuremath {353{,}929.06}&\pgfutilensuremath {637{,}072.31}&\pgfutilensuremath {1{,}486{,}502.06}&--&$\mathbf{
  \pgfmathprintnumber[precision=2,zerofill,fixed] {85648.12}}$&\pgfutilensuremath {97{,}859.81}\\%
Web-UK&\pgfutilensuremath {533{,}427.01}&\pgfutilensuremath {616{,}486.63}&\pgfutilensuremath {635{,}946.11}&\pgfutilensuremath {644{,}624.98}&\pgfutilensuremath {633{,}044.02}&\pgfutilensuremath {647{,}050.45}&\pgfutilensuremath {659{,}961.81}&\pgfutilensuremath {669{,}742.03}&\pgfutilensuremath {444{,}407.11}&\pgfutilensuremath {799{,}932.80}&\pgfutilensuremath {1{,}866{,}509.86}&--&--&$\mathbf{
  \pgfmathprintnumber[precision=2,zerofill,fixed] {266342.83}}$\\\bottomrule %
\end {tabular}%
}
}
\vspace*{.5cm}
 \subfloat[Construction Time (ms)]{
 \resizebox{\textwidth}{!}{
\begin {tabular}{rrrrrrrrrrrrrrr}%
\toprule 
& \multicolumn{4}{c}{\bfseries Ferrari-L} & \multicolumn{4}{c}{\bfseries
  Ferrari-G} & \multicolumn{4}{c}{\bfseries Grail} 
& 
& 
\\
\cmidrule(r){2-5} \cmidrule(r){6-9} \cmidrule(r){10-13}
\bfseries Dataset & 
\multicolumn{1}{c}{\footnotesize $k=1$} & \multicolumn{1}{c}{\footnotesize $k=2$} & \multicolumn{1}{c}{\footnotesize $k=5$} & \multicolumn{1}{c}{\footnotesize $k=10$} &
\multicolumn{1}{c}{\footnotesize $k=1$} & \multicolumn{1}{c}{\footnotesize $k=2$} & \multicolumn{1}{c}{\footnotesize $k=5$} & \multicolumn{1}{c}{\footnotesize $k=10$} &
\multicolumn{1}{c}{\footnotesize $d=1$} & \multicolumn{1}{c}{\footnotesize $d=2$} & \multicolumn{1}{c}{\footnotesize $d=5$} & \multicolumn{1}{c}{\footnotesize $d=10$} &
\bfseries Interval & \bfseries PWAH-8 \\
\midrule
YAGO2&\pgfutilensuremath {27{,}475.10}&\pgfutilensuremath {27{,}713.50}&\pgfutilensuremath {27{,}679.40}&\pgfutilensuremath {26{,}778.80}&\pgfutilensuremath {28{,}169.40}&\pgfutilensuremath {26{,}865.30}&\pgfutilensuremath {26{,}614.70}&\pgfutilensuremath {26{,}511.40}&\pgfutilensuremath {9{,}046.63}&\pgfutilensuremath {17{,}163.00}&\pgfutilensuremath {59{,}587.00}&\pgfutilensuremath {123{,}328.00}&$\mathbf{
  \pgfmathprintnumber[precision=2,zerofill,fixed] {5844.87}}$&\pgfutilensuremath {9{,}236.71}\\%
GovWild&\pgfutilensuremath {12{,}025.40}&\pgfutilensuremath {12{,}998.80}&\pgfutilensuremath {15{,}824.30}&\pgfutilensuremath {16{,}565.50}&\pgfutilensuremath {15{,}487.20}&\pgfutilensuremath {18{,}045.30}&\pgfutilensuremath {14{,}874.80}&\pgfutilensuremath {14{,}941.60}&$\mathbf{
  \pgfmathprintnumber[precision=2,zerofill,fixed] {3535.81}}$&\pgfutilensuremath {6{,}756.67}&\pgfutilensuremath {24{,}482.30}&\pgfutilensuremath {50{,}129.00}&\pgfutilensuremath {15{,}060.55}&\pgfutilensuremath {20{,}703.06}\\%
Twitter&\pgfutilensuremath {13{,}434.70}&\pgfutilensuremath {13{,}065.40}&\pgfutilensuremath {13{,}403.20}&\pgfutilensuremath {13{,}768.00}&\pgfutilensuremath {13{,}442.50}&\pgfutilensuremath {13{,}897.20}&\pgfutilensuremath {13{,}877.70}&\pgfutilensuremath {13{,}885.50}&$\mathbf{
  \pgfmathprintnumber[precision=2,zerofill,fixed] {5784.66}}$&\pgfutilensuremath {9{,}717.39}&\pgfutilensuremath {70{,}038.80}&\pgfutilensuremath {161{,}704.00}&\pgfutilensuremath {36{,}480.57}&\pgfutilensuremath {8{,}219.09}\\%
Web-UK&\pgfutilensuremath {18{,}145.00}&\pgfutilensuremath {17{,}604.90}&\pgfutilensuremath {18{,}029.70}&\pgfutilensuremath {18{,}506.30}&\pgfutilensuremath {18{,}463.30}&\pgfutilensuremath {18{,}754.40}&\pgfutilensuremath {19{,}056.40}&\pgfutilensuremath {19{,}202.70}&$\mathbf{
  \pgfmathprintnumber[precision=2,zerofill,fixed] {7273.53}}$&\pgfutilensuremath {12{,}275.90}&\pgfutilensuremath {82{,}185.90}&\pgfutilensuremath {181{,}947.00}&--&\pgfutilensuremath {166{,}531.10}\\\bottomrule %
\end {tabular}%
}
}
\caption{Experimental Evaluation on Web Datasets -- Indexing (Varying Size Constraint)}
\label{tab:exp-web-index-vark}
\end{sidewaystable}

\bibliographystyle{plain}
\bibliography{ferrari}
\end{document}